\documentclass[11pt]{article}

\usepackage{amsmath,amsthm,amssymb,mathtools}
\usepackage[T1]{fontenc}
\usepackage{lmodern}
\usepackage{microtype}
\usepackage[margin=1.15in]{geometry}
\usepackage{graphicx}
\graphicspath{{./Image/}}
\usepackage{booktabs}
\usepackage{array}
\usepackage{longtable}
\usepackage{cite}
\usepackage[hidelinks]{hyperref}
\usepackage{titlesec}
\usepackage{xcolor}

\titleformat{\section}[block]{\normalfont\large\bfseries}{\thesection.}{0.6em}{}
\titleformat{\subsection}[block]{\normalfont\normalsize\bfseries}{\thesubsection.}{0.5em}{}
\titleformat{\subsubsection}[block]{\normalfont\normalsize\itshape}{\thesubsubsection.}{0.5em}{}
\titlespacing*{\section}{0pt}{1.4\baselineskip}{0.7\baselineskip}
\titlespacing*{\subsection}{0pt}{1.0\baselineskip}{0.5\baselineskip}

\theoremstyle{plain}
\newtheorem{lemma}{Lemma}
\newtheorem{proposition}{Proposition}
\newtheorem{corollary}{Corollary}
\theoremstyle{definition}
\newtheorem{definition}{Definition}
\newtheorem{assumption}{Assumption}

\newcommand{\setcor}[1]{\renewcommand{\thecorollary}{#1}}

\newcommand{\R}{\mathbb{R}}

\newcommand{\din}{d^{\mathrm{in}}}
\newcommand{\dout}{d^{\mathrm{out}}}
\newcommand{\trace}{\operatorname{tr}}

\newcommand{\nullity}{\operatorname{null}}
\newcommand{\Real}{\operatorname{Re}}
\DeclareMathOperator{\spec}{spec}
\numberwithin{equation}{section}
\usepackage{setspace}

\title{\bfseries Assortativity and the Distribution of Spectral Gaps in Directed Networks}
\author{Sreerag Puravankara\thanks{Indian Institute of Management Kozhikode,
Kozhikode 673570, Kerala, India.}
\and Vipin P. Veetil\footnotemark[1]}
\date{}

\begin{document}
\maketitle
\setstretch{1.3}

\begin{abstract}
This paper studies how the distribution of spectral gaps of a weighted directed network varies with its
assortativity. We prove that the sum of the distances of each of the eigenvalues from the Perron root along the real axis, which we call the `spectral depth', decreases as the network is rewired to be less assortative. The result is conditional on the network's Perron vectors ordering the rewired vertices in the same way as their strengths do. We also develop a method to determine whether a particular spectral gap contracts or expands under a rewiring that changes the network's assortativity. It turns out that the direction of any one gap depends on far more than the network's assortativity. In computational experiments on heavy-tailed networks, disassortative rewiring lowers the spectral depth by about two fifths, contracts most strongly the gaps in the middle of the spectrum, and collapses a growing part of the spectrum to zero. The experiments also show that a sequence of disassortative rewirings tends to narrow the first spectral gap, while a sequence of assortative rewirings widens it.

\end{abstract}

\medskip


\newpage
\section{Introduction}\label{sec:intro}

The gaps between the eigenvalues of a network set the time scales of linear dynamics on it. The
first spectral gap, between the leading eigenvalue and the second, sets the rate at which the network as
a whole converges to its long-run state
\cite{levinperes}. The spectral gaps further down set the rates at which parts of the network converge
within themselves. A network composed of prominent community structures that are only loosely tied to one another will tend to exhibit faster convergence `within' each community, possibly to different states, than for the network as a whole \cite{simonando}. This information is embedded in the distribution of spectral gaps. More specifically, a network of loosely tied communities has one leading eigenvalue for each
community, with the leading eigenvalues of the communities sitting close together, separated from the rest of the spectrum by a wide gap
\cite{leegharan,arenas}. The narrow gaps among the leading eigenvalues set the slow convergence of the communities to one
another, and the wide gap sets the fast convergence within each community.\footnote{The
cluster of leading eigenvalues arises most directly when the weight matrix is normalized so that
every row or every column sums to one. Without that normalization, the cluster arises when the communities
have leading eigenvalues of comparable size. In a directed network the rates that the gaps set govern the long run, and the transient that precedes it can last longer when the eigenvectors are far from orthogonal \cite{trefethenembree}.} In short, the distribution of spectral gaps is a matter of great interest when it comes to questions of dynamics on networks, whether the dynamics be economic, epidemiological, social or informational.

It so happens that most networks of practical interest are either assortative or disassortative,
where by assortativity we mean how a network pairs its vertices by degree, or by strength when
its edges carry weights. We say that a network is
assortative if it joins its vertices of high degree to one another and its vertices of low degree
to one another, and disassortative if it joins vertices of high degree to vertices of low degree.
Social networks tend to be assortative, whereas the Internet, protein-interaction networks and
several other biological and technological networks are typically disassortative
\cite{newman2002,newman2003,pastorsatorras,maslov}. Firm buyer-seller networks too are mildly
disassortative \cite{bacilieri}. Such assortativity could, in principle, shape the dynamics on
these networks if it influences the distribution of spectral gaps. The question of how assortativity influences spectral gaps is therefore of practical significance in a variety of sciences. Previous work gives us an inkling of how the distribution of spectral gaps changes
with assortativity, but little of it rests on the bedrock of theory. Which is precisely what we attempt to provide in
this paper.

We work with directed networks whose edges carry weights and whose eigenvalues need not be real. We change assortativity by
moving weight between edges while holding the in-strength and out-strength of every vertex fixed.
More specifically, the rewiring takes weight from two edges and places it on the two edges
obtained by exchanging their targets. We measure assortativity by the correlation between the
in-strength at the source of an edge and the out-strength at its target \cite{foster}.\footnote{We use this
pairing because the response of the Perron root to a rewiring multiplies a difference of the left
Perron vector at the two sources by a difference of the right Perron vector at the two targets, and
the left Perron vector follows in-strength while the right one follows out-strength. Like Newman's
coefficient, this correlation is a Pearson coefficient. The lower bound of Newman's coefficient
depends on the size of the network and approaches zero for heavy-tailed degree sequences
\cite{litvak}. Our analytical results use only the sign of the change in the correlation, which such
a bound does not affect.}
The weight transfer is modeled as a continuous process, which we trace from the original network to
the one arrived at through the rewiring. The evolution of the eigenvalues along the transfer can
then be characterized using simple but powerful perturbation theory \cite{kato,stewartsun,greenbaum}.

We begin with the rather innocuous observation that a rewiring that neither adds nor removes
self-loops preserves the trace of the weight matrix. Note that our rewiring process  neither adds nor removes a self-loop since the four vertices it involves are distinct. This means that our rewiring process does not alter the sum of
the eigenvalues, which is of course the trace. Which in turn implies that if the Perron root
changes, the other eigenvalues must on balance move the other way to preserve the sum. We call the total distance
of the other eigenvalues beneath the Perron root, measured along the real axis, the ``spectral
depth''. Consider dynamics in
continuous time in which the state of each vertex changes at a rate equal to the weighted sum of
the states of the vertices it links to. The state of the network is then a sum of modes, one for
each eigenvalue. Relative to the leading mode, which belongs to the Perron root, each other mode
decays at a rate equal to the distance of its eigenvalue beneath the Perron root along the real axis.
The spectral depth is the sum of these rates. When the spectrum is real, each eigenvalue lies
beneath the Perron root by the sum of the gaps between them. So the spectral depth is a weighted sum of all the gaps. The spectral depth changes by exactly the number of vertices times the change in the Perron root. Elementary as it is, this tie between
the spectral depth and the Perron root makes every result on the Perron root a result on the gaps
taken together.

The Perron root sets how fast the number of walks on a network grows with their length, and
it responds most to weight on edges that lead from vertices at which many walks arrive to vertices
from which many walks depart. When the left and right Perron vectors of a network follow its strengths, the edges to which the
Perron root responds most lead from vertices of high in-strength to vertices of high out-strength. A
disassortative rewiring moves weight off such an edge, and off an edge between vertices of low
strength, onto two edges that join vertices of high strength to vertices of low strength.

Whenever the left Perron vector ranks the two sources of the rewiring in the order of their
in-strengths and the right Perron vector ranks the two targets in the order of their out-strengths,
a disassortative rewiring lowers the Perron root. In earlier work we had bounded how far each Perron vector lies from its strength vector over all
vertices \cite{pv}. In this paper we extend our earlier work with a condition for the ranking, which
limits how far the Perron vectors may lie from the strength vectors at the four rewired vertices
alone.
The spectral depth falls with the Perron root. As the Perron root falls, the other modes of the network taken together decay more slowly relative to
its leading mode. When the spectrum is real, a falling spectral depth leaves room for some gaps to widen while others narrow by more. But the
spectral depth sets an upper bound on every gap and on the share of gaps above any given size, and
neither upper bound can rise as the spectral depth falls. All of which is to say that a change in assortativity moves the
gaps taken together through one property, the Perron root. As the Perron root falls, the gaps taken
together narrow.

What remains to be established is how a particular gap responds to a change in assortativity. We
derive an exact criterion for the direction of each gap along a rewiring.\footnote{An ordered gap
needs a real spectrum, and a directed network can have eigenvalues that are not real, so for a pair
of them we follow the distance between the two in place of a gap. This distance moves with the
velocities of the two eigenvalues, as a gap does.} A gap moves with the velocities of the two eigenvalues that bound it. The gap contracts exactly
when the upper eigenvalue has the smaller velocity. When the eigenvalues are distinct, the velocity of each is a weighted sum of the counts
of walks from the two targets of the rewiring back to its two sources. The sum runs over all
lengths less than the number of vertices, and its weights depend on every eigenvalue. The upper
bounds on the gaps rest on a single summary statistic, the spectral depth, but the criterion rests on
none. We show that the counts of all but the longest of these walks are compatible both with the
contraction and with the expansion of any gap. When the counts of the shortest of these walks vanish
and the count of the next longer ones is nonnegative, a falling Perron root does force some of the
gaps to alternate between contraction and expansion. The alternation runs down the spectrum and
begins with a contracting gap. But there is no general way, at least none that we have found, of
saying how a gap responds to a change in assortativity without knowing a fair bit about the network.
The exact criterion needs every eigenvalue of the network, and it becomes unstable when eigenvalues
cluster. Therefore, we also give approximations to it that use only a group of leading eigenvalues
\cite{shuman,trefethen,saad}. The direction of a gap at the start of a rewiring may reverse along
the way. We give a sufficient condition, checked at finitely many points along the path, under which a gap contracts at every point of the rewiring.

Though our analytical results sign the effect of assortativity
on the spectral gaps, they do not provide a measure of its size. We use computational experiments to measure the size of the effect. More specifically, we implement our rewiring algorithm on simulated networks with heavy-tailed
degrees and follow the spectrum through long sequences of rewirings. Each rewiring transfers as much
weight as the two edges it draws from allow. We select rewirings by the sign of their effect on
assortativity alone. On large directed networks,
disassortative rewiring lowers the spectral depth substantially. On symmetric networks, whose
spectra are real, disassortative rewiring contracts most strongly the gaps in the middle of the
spectrum. Assortative rewiring contracts most strongly the gaps just inside the two ends of the
spectrum. A disassortative rewiring contracts the first gap more often than not, and an assortative rewiring
less often than not, but neither does so nearly always. Along a long sequence of rewirings of one sign, the first gap moves the same way on
nearly every network. It falls under disassortative rewiring and rises under assortative rewiring.
On heavy-tailed networks without self-loops, disassortative rewiring also raises the
share of gaps equal to zero. Put simply, a growing part of the spectrum collapses onto zero.

\subsection{Related literature}

Mathematicians before us have studied how eigenvalues move with assortativity, but they have said
little about how the distribution of spectral gaps moves with it. What has been proved concerns mainly the Perron root, the largest eigenvalue of the adjacency
matrix.\footnote{Since degree-preserving rewiring leaves the numbers of walks of lengths one and two
unchanged, Newman's coefficient varies only through the number of walks of length three, with which
it rises. The ratio of the number of walks of length three to the number of walks of length two is a
lower bound on the Perron root, and with its denominator fixed, the bound rises with assortativity.
In simulations the Perron root itself rises with assortativity as well \cite{vanmieghem}. The
mean-field approximation of the Perron root, the ratio of the second moment of the degrees to the
first, gains a correction factor that rises with the degree correlation \cite{restrepo}.
B{\i}y{\i}ko\u{g}lu and Leydold \cite{biyikoglu} describe the graphs that maximize the Perron root
at a fixed degree sequence, Atay and B{\i}y{\i}ko\u{g}lu \cite{atay} show them to be strongly
assortative, with degree and eigenvector centralities that coincide, and Chung, Lu and Vu
\cite{chunglu} give the asymptotic value of the Perron root on random graphs with given expected
degrees. Noldus and Van Mieghem \cite{noldus} survey the measurement of assortativity and its
effects on the spectrum.} And what is known about spectral gaps comes from computational experiments, by and large on binary networks (in which every edge carries the same weight).\footnote{The one
partial exception is the approximation of the Perron root by Restrepo, Ott and Hunt \cite{restrepo},
whose derivation allows a general nonnegative weight matrix, though their numerical tests are on
binary networks.} For instance, in simulations of power-law networks with degree assortativity or disassortativity, \cite{silva} find that the first spectral gap, the distance between the two largest eigenvalues, is narrower
when the network is disassortative and wider when it is assortative.  Previous work has also shown that assortativity moves the gaps through the density of eigenvalues.
Rewiring to a prescribed degree correlation changes that density markedly, and with it the mean gap
at each point of the spectrum. The distribution of a gap relative to the mean gap around it does not however
change \cite{jalan}.\footnote{Homophily, which pairs vertices by
type rather than by degree, narrows the first gap of the interaction matrix in a model of learning on
a network, and so slows convergence to consensus, whatever the degree distribution
\cite{golubjackson}.} What distinguishes this paper is that we develop analytical results for directed and weighted networks, using a strength-preserving rewiring, the weighted and directed form of the degree-preserving
rewiring with which Van Mieghem et al. \cite{vanmieghem} and Noldus and Van Mieghem \cite{noldus}
change assortativity.

\subsection{Organization of the paper}

Section~\ref{sec:switch} defines the rewiring, assortativity and the gap distribution.
Section~\ref{sec:whole} ties the spectral depth to the Perron root, gives a sufficient condition
under which a disassortative rewiring lowers the Perron root, and draws the consequences for the gap distribution. Section~\ref{sec:individual} derives the exact
criterion for whether a single gap contracts or expands, along with approximations to it, and shows
when a falling Perron root forces the gaps to alternate between contraction and expansion. The section also follows pairs of complex eigenvalues and ends with a sufficient
condition under which a gap contracts at every point of a rewiring. Section~\ref{sec:computations} reports
the computational experiments, and Section~\ref{sec:conclusion} concludes.
Appendix~\ref{app:lemmas} collects the auxiliary lemmas, Appendix~\ref{app:numerics} gives the
details of the computational experiments, and Appendix~\ref{app:notation} lists the notation.

\section{The rewiring, assortativity and the gaps}\label{sec:switch}

We write \(e_i\) for the \(i\)-th standard basis vector, \(E_{ij}=e_ie_j^{\mathsf T}\) for the
matrix whose only nonzero entry is a one in position \((i,j)\), and \(\mathbf{1}\) for the vector
of ones. The transpose of a matrix \(X\) is \(X^{\mathsf T}\), its trace is \(\trace X\), its
spectrum is \(\spec X\), and the dimension of its null space is \(\nullity X\). The diagonal
matrix with the entries of a vector \(v\) on its diagonal is \(\operatorname{diag}(v)\), and
\(\operatorname{diag}X\) is the vector of diagonal entries of \(X\). We use the Euclidean norm \(\lVert v\rVert_2\)
of a vector and the spectral norm \(\lVert X\rVert_2\) of a matrix. The unit point mass at \(u\)
is \(\delta_u\), and \(\delta\) without a subscript denotes the weight that a rewiring transfers.
Table~\ref{tab:notation} in Appendix~\ref{app:notation} lists the notation of the paper.

\subsection{Strength-preserving rewiring}

Let \(W=(w_{ij})\in\R^{n\times n}\) be the weight matrix of a directed network on \(n\)
vertices, in which \(w_{ij}>0\) is the weight of the edge from vertex \(i\) to vertex \(j\) and
\(w_{ij}=0\) when there is no such edge. Fix four distinct vertices \(a,b,c,d\). The
directed two-switch removes weight from the donor edges \(a\to b\) and \(c\to d\)
and transfers it to the crossed edges \(a\to d\) and \(c\to b\).

\begin{definition}[Transfer path]\label{def:switch}
Put \(x=e_a-e_c\), \(y=e_d-e_b\) and \(B=xy^{\mathsf T}\), and let
\(0<\delta\leq\min\{w_{ab},w_{cd}\}\). The transfer path of the switch with transferred
weight \(\delta\) is
\begin{equation}
W(t)=W+t\delta B,
\qquad
0\leq t\leq1.
\label{eq:path}
\end{equation}
\end{definition}

Here \(\delta\) is the total weight transferred, and \(t\) is the fraction of it
moved so far. The bound on \(\delta\) in Definition~\ref{def:switch}
requires the two donor edges to be present, so that \(w_{ab}>0\) and \(w_{cd}>0\). The same bound keeps
every entry of \(W(t)\) nonnegative.
We call a switch whose donor edges are present admissible. We call an admissible switch together
with the transfer of weight along its path a rewiring. The path runs from the original matrix \(W(0)=W\) to the rewired matrix \(W(1)\). A completed
transfer is the endpoint \(t=1\) of the path, at which the whole weight \(\delta\) has moved, and a
partial transfer is any earlier point.

For a weighted network every intermediate matrix \(W(t)\) is itself a weighted
network, so the partial transfer up to \(t\) is itself a completed transfer, of the weight \(t\delta\). For a binary
network with \(w_{ad}=w_{cb}=0\) and \(\delta=1\), the intermediate matrices
are weighted networks between the original network and the switched one, and the endpoint
\(W(1)\) is the result of the usual unweighted two-switch. For a symmetric network we apply the
switch together with its transpose along the path \(W(t)=W+t\delta(xy^{\mathsf T}+yx^{\mathsf T})\),
so that \(W(t)\) stays symmetric. We call the transfer along this path the symmetric transfer.

The rewiring changes how the network pairs its vertices without changing how much weight any
vertex sends or receives. The path has constant velocity \(W'(t)=\delta B\), with
\(B=E_{ad}+E_{cb}-E_{ab}-E_{cd}\). The switch alters four entries of \(W\) and
no others. The sources \(a\) and \(c\) each lose weight on one outgoing edge and gain the same
weight on another outgoing edge. The targets \(b\) and \(d\) each lose weight on one incoming edge
and gain the same weight on another incoming edge. So every vertex keeps its out-strength and
its in-strength throughout the transfer. And because \(a,b,c,d\) are distinct the switch leaves
the diagonal alone, so the trace of the weight matrix does not change. The symmetric transfer,
whose velocity is \(\delta(B+B^{\mathsf T})\), preserves the strengths and the trace in the
same way (Lemma~\ref{lem:preserve}).

The spectral radius of the nonnegative matrix \(W(t)\) is itself an eigenvalue of \(W(t)\), and
the largest real one \cite{hornjohnson}. We call it the Perron root \(\rho(t)\). Statements that use
the Perron vectors of \(W(t)\) or the derivative of the Perron root along the path rest on the
irreducibility of \(W(t)\), which we now state as an assumption.

\begin{assumption}[Irreducibility]\label{ass:pf}
The matrix \(W(t)\) is irreducible for every
\(t\in[0,1]\).
\end{assumption}

The assumption says that every vertex can reach every other vertex along edges of
positive weight at every point of the path. Which means that the rewiring never splits the
network into parts that cannot reach one another. For \(0<t<1\) every edge of positive weight in
\(W(0)\) or in \(W(1)\) also has positive weight in \(W(t)\), so the assumption holds exactly when
\(W(0)\) and \(W(1)\) are irreducible. Under Assumption~\ref{ass:pf}, the Perron root is a simple
eigenvalue and its left and right Perron vectors are positive \cite{hornjohnson}. The Perron root
and vectors can also be chosen analytically along the transfer path \cite{kato}.

\subsection{Assortativity}

Write \(\dout=W\mathbf{1}\) and \(\din=W^{\mathsf T}\mathbf{1}\) for the
out-strength and in-strength vectors of \(W\). Write \(\omega=\sum_{ij}w_{ij}\) for the total weight
of the network and \(\pi_{ij}=w_{ij}/\omega\) for the share of that weight carried by the edge
\(i\to j\). We regard each edge, weighted by \(\pi_{ij}\), as carrying the in-strength of its source
and the out-strength of its target. The source-in/target-out covariance of \(W\) is
\begin{equation}
C_{io}(W)
=
\sum_{ij}
\pi_{ij}
\bigl(\din_i-\bar d_{\mathrm{in}}\bigr)
\bigl(\dout_j-\bar d_{\mathrm{out}}\bigr),
\qquad
\bar d_{\mathrm{in}}
=
\sum_{ij}\pi_{ij}\din_i,
\qquad
\bar d_{\mathrm{out}}
=
\sum_{ij}\pi_{ij}\dout_j.
\label{eq:Cio}
\end{equation}

Along the path \eqref{eq:path}, the strength vectors, the marginals of \(\pi\), the two means, and
the variances of source in-strength and target out-strength remain constant. Only the weights
\(\pi_{ij}\), which pair sources with targets, change. Lemma~\ref{lem:assort} gives
\begin{equation}
\frac{d}{dt}C_{io}\bigl(W(t)\bigr)
=
\frac{\delta}{\omega}
\bigl(\din_a-\din_c\bigr)
\bigl(\dout_d-\dout_b\bigr).
\label{eq:assortderiv}
\end{equation}

Because the right-hand side of \eqref{eq:assortderiv} is constant in \(t\), the
change \(C_{io}(W(1))-C_{io}(W(0))\) over a completed transfer equals that
right-hand side, and so is proportional to the transferred weight \(\delta\).
We call the rewiring disassortative when the product on the right-hand side
of \eqref{eq:assortderiv} is negative, and \(C_{io}\) then decreases throughout
the transfer. When the product is positive we call the rewiring assortative,
and \(C_{io}\) increases throughout the transfer. When the product is zero, \(C_{io}\) does not
change. These names always refer to the pairing of the
in-strength of a source with the out-strength of its target.\footnote{A directed edge admits four pairings of
in-strengths and out-strengths at its two ends. We treat only the
source-in/target-out pairing of \eqref{eq:assortderiv}. The other three, source-out/target-in,
source-in/target-in and source-out/target-out, would
need a corresponding relation between strengths and Perron coordinates.}
Under the symmetric transfer, the derivative is twice the right-hand side
of \eqref{eq:assortderiv} (Lemma~\ref{lem:assort}), and we name the rewiring by the sign of the
same product. In what follows we concentrate on disassortative rewirings. All our analytical results carry over to
assortative rewirings as well.

\subsection{The gap distribution}

A directed matrix whose eigenvalues are not all real has no natural ordering of its eigenvalues, so
we define the gaps only for a real spectrum, and for any other spectrum we follow the distance
between selected pairs of eigenvalues instead.

\begin{definition}[Gaps and the gap distribution]\label{def:gaps}
Let the spectrum of \(W\) be real and ordered as
\(\lambda_1\geq\lambda_2\geq\cdots\geq\lambda_n\), with every eigenvalue repeated
according to its algebraic multiplicity. The adjacent gaps of \(W\) are
\begin{equation}
g_i=\lambda_i-\lambda_{i+1}\geq0,
\qquad
i=1,\dots,n-1.
\label{eq:gaps}
\end{equation}
The gap distribution of \(W\) is
\[
\nu_W
=
\frac{1}{n-1}\sum_{i=1}^{n-1}\delta_{g_i}.
\]
\end{definition}

Because the eigenvalues are reordered at each matrix, the gap distribution
\(\nu_W\) does not depend on how the eigenvalues, followed continuously along a path as branches, are
labeled, even where two branches cross. Its support lies in \([0,\infty)\), and coincident eigenvalues
contribute gaps of length zero.

\section{Assortativity and the gap distribution}\label{sec:whole}

When the Perron root changes, the rest of the spectrum must move too, for the
eigenvalues sum to the trace and the rewiring leaves the trace alone. So a
fall in the Perron root has to be made up by a rise, on balance, in the other
eigenvalues. Put differently, the total distance of the other eigenvalues
beneath the Perron root, measured along the real axis, falls by exactly $n$
times the fall in the Perron root. We call this total distance the `spectral depth'.

\begin{definition}[Spectral depth]\label{def:depth}
The spectral depth of a weight matrix $W$ with Perron root $\rho$ is
\begin{equation}
\mathcal{R}(W)
=
n\rho-\trace W.
\label{eq:R}
\end{equation}
\end{definition}

The eigenvalues $\lambda_1=\rho,\lambda_2,\dots,\lambda_n$ of $W$, repeated according to their
algebraic multiplicities, sum to the trace, so $\mathcal{R}(W)=\sum_{r=2}^{n}(\rho-\Real\lambda_r)$.
Each term of this sum is the distance of an eigenvalue beneath the Perron root along the real axis,
which is never negative because $\Real\lambda_r\le\lvert\lambda_r\rvert\le\rho$. The spectral depth
is real whatever the spectrum, for $\rho$ and $\trace W$ are real.

\subsection{The spectral depth and the Perron root}

Suppose that the spectrum of $W(t)$ is real and ordered, so that
$\lambda_1(t)=\rho(t)\ge\lambda_2(t)\ge\cdots\ge\lambda_n(t)$, and let
$g_i(t)=\lambda_i(t)-\lambda_{i+1}(t)$ be its gaps. Each eigenvalue then lies beneath the Perron
root by the sum of the gaps above it, so the spectral depth of a real spectrum is a weighted sum
of its gaps.

\begin{proposition}[Spectral depth and the Perron root]\label{prop:global}
Every rewiring satisfies
\begin{equation}
\mathcal{R}\bigl(W(1)\bigr)-\mathcal{R}\bigl(W(0)\bigr)
=
n\bigl(\rho(1)-\rho(0)\bigr).
\label{eq:globallaw}
\end{equation}
Under Assumption~\ref{ass:pf}, the rates of change at each point of the path satisfy
\begin{equation}
\frac{d}{dt}\mathcal{R}\bigl(W(t)\bigr)
=
n\rho'(t),
\qquad t\in[0,1].
\label{eq:Rderivative}
\end{equation}
If the spectrum of $W(t)$ is real and ordered, then
\begin{equation}
\mathcal{R}\bigl(W(t)\bigr)
=
\sum_{i=1}^{n-1}(n-i)g_i(t)
=
\sum_{r=2}^{n}\bigl(\rho(t)-\lambda_r(t)\bigr)
=
n\rho(t)-\trace W(t).
\label{eq:Rtrace}
\end{equation}
In that case the gap distribution also satisfies,
for every $u>0$,
\begin{equation}
\nu_{W(t)}\bigl([u,\infty)\bigr)
\le
\frac{1}{n-1}
\left\lfloor
\frac{\sqrt{1+8\mathcal{R}(W(t))/u}-1}{2}
\right\rfloor .
\label{eq:tail}
\end{equation}
\end{proposition}

\begin{proof}
Along a rewiring,
$\mathcal{R}(W(1))-\mathcal{R}(W(0))=n(\rho(1)-\rho(0))-(\trace W(1)-\trace W(0))$.
The rewiring modifies only off-diagonal entries because $a,b,c,d$ are distinct,
and Lemma~\ref{lem:preserve} gives $\trace W(1)=\trace W(0)$. This proves
\eqref{eq:globallaw}. Under Assumption~\ref{ass:pf}, the Perron root remains simple and
differentiable along the path. The trace is constant, so differentiating $n\rho(t)-\trace W(t)$
gives \eqref{eq:Rderivative}.

For a real ordered spectrum, the separation of the $r$-th eigenvalue from the
Perron root is $\rho-\lambda_r=g_1+g_2+\cdots+g_{r-1}$. When these separations are summed over
$r=2,\ldots,n$, the gap $g_i$ appears $n-i$ times, once for each of the eigenvalues
$\lambda_{i+1},\ldots,\lambda_n$. So $\sum_{r=2}^{n}(\rho-\lambda_r)=\sum_{i=1}^{n-1}(n-i)g_i$. Counted
with algebraic multiplicity, the eigenvalues sum to the trace, and
\[
\sum_{r=2}^{n}(\rho-\lambda_r)
=
(n-1)\rho-\sum_{r=2}^{n}\lambda_r
=
n\rho-\sum_{r=1}^{n}\lambda_r
=
n\rho-\trace W .
\]

Finally, fix $u>0$ and let $k$ be the number of gaps that are at least $u$.
Their weights $n-i$ are $k$ distinct positive integers, whose sum is at least
$1+2+\cdots+k=k(k+1)/2$. Because the ordered gaps are nonnegative,
$\mathcal{R}(W(t))\ge u\,k(k+1)/2$, and solving for the integer $k$ gives
\[
k
\le
\left\lfloor
\frac{\sqrt{1+8\mathcal{R}(W(t))/u}-1}{2}
\right\rfloor .
\]
The share of gaps that are at least $u$ is $\nu_{W(t)}([u,\infty))=k/(n-1)$, so dividing the
bound on $k$ by $n-1$ gives \eqref{eq:tail}.
\end{proof}

By Proposition~\ref{prop:global}, every result in the literature on the Perron root is also a
result on the spectral gaps. The identity \eqref{eq:globallaw} compares
the rewired matrix $W(1)$ directly with the original matrix $W(0)$. It holds for every rewiring and
needs only the Perron roots of these two matrices.\footnote{The Perron root of a stochastic network cannot move at all. Its weight
matrix $\mathcal{P}=\operatorname{diag}(\dout)^{-1}W$ divides each row of $W$ by its out-strength, which
Assumption~\ref{ass:pf} makes positive, so every row of $\mathcal{P}$ sums to one and its Perron root is one
whatever the rewiring. A rewiring keeps every out-strength and leaves the diagonal alone, so the
trace of $\mathcal{P}(t)=\operatorname{diag}(\dout)^{-1}W(t)$ is constant and the spectral depth
$\mathcal{R}(\mathcal{P}(t))=n-\trace\mathcal{P}$ does not change. All of which is to say, in a stochastic network all
assortativity can do is redistribute the spectrum. On a symmetric network under the symmetric
transfer, $\mathcal{P}(t)$ is similar to the symmetric matrix
$\operatorname{diag}(\dout)^{-1/2}W(t)\operatorname{diag}(\dout)^{-1/2}$. So its spectrum is real,
and a rewiring that narrows the first gap must widen the gaps beneath it by the same weighted amount.
In continuous time the modes decay toward the long-run state at the rates $1-\lambda_r(\mathcal{P})$, and
slowing some modes speeds up other modes by the same total amount.} By \eqref{eq:globallaw}, a rewiring that preserves strengths and leaves the diagonal alone changes
the spectral depth by $n$ times whatever it does to the Perron root, so anything known about the
change in the Perron root carries over to the spectral depth, and no eigenvector needs to be
computed. For a real ordered spectrum, the change in the spectral depth is also the change in the
weighted sum $\sum_i(n-i)g_i$ of the gaps. The identity \eqref{eq:Rderivative} gives the rate at
which the spectral depth changes at each point of the path, and it needs Assumption~\ref{ass:pf}.

For a symmetric network, Newman's coefficient can be written in terms of the
numbers of walks of lengths one to three. The ratio of the number of walks of
length three to the number of walks of length two bounds the Perron root from below, and the ratio rises with the coefficient at a fixed degree sequence. In
simulation the Perron root itself rises with the coefficient \cite{vanmieghem}.
By Proposition~\ref{prop:global}, multiplying this lower bound on the Perron root by $n$ and
subtracting the trace gives a lower bound on the spectral depth of a real spectrum that rises with assortativity. And the
rise of the Perron root seen in simulation implies a rise in the spectral depth $n$ times as large.

\subsection{The response of the Perron root}\label{sec:perron}

Whether the spectral depth rises or falls depends on how
the Perron root responds to the rewiring. Under Assumption~\ref{ass:pf} the Perron root is simple,
with unit positive left and right Perron vectors \(\ell(t)\) and \(r(t)\). Lemma~\ref{lem:perron}
gives its exact response to the rewiring,
\[
\rho'(t)
=
\delta\,
\frac{
\bigl(\ell_a(t)-\ell_c(t)\bigr)
\bigl(r_d(t)-r_b(t)\bigr)
}{
\ell(t)^{\mathsf T}r(t)
}.
\]
The sign of \(\rho'(t)\) is the sign of a product of two
differences of Perron coordinates, one between the sources \(a\) and \(c\) on
the left Perron vector and one between the targets \(d\) and \(b\) on the
right Perron vector. The sign of the change in assortativity is the sign of
the product of the corresponding differences of strengths in
\eqref{eq:assortderiv}. The Perron vectors are unit vectors, so we compare
them with the strength vectors normalized in the same way.

\begin{definition}[Strength contrasts and coordinate errors]\label{def:contrast}
Let \(p=\din/\lVert\din\rVert_2\) and \(q=\dout/\lVert\dout\rVert_2\) be the normalized
in-strength and out-strength vectors. Define the source,
target and combined strength contrasts by
\begin{equation}
\chi_{\mathrm{in}}=p_a-p_c,
\qquad
\chi_{\mathrm{out}}=q_d-q_b,
\qquad
\chi_{io}=\chi_{\mathrm{in}}\chi_{\mathrm{out}}.
\label{eq:chi}
\end{equation}
Under Assumption~\ref{ass:pf}, the corresponding coordinate errors are
\begin{equation}
\begin{aligned}
\varepsilon_{\mathrm{in}}(t)
&=
\bigl(\ell_a(t)-p_a\bigr)
-
\bigl(\ell_c(t)-p_c\bigr),\\
\varepsilon_{\mathrm{out}}(t)
&=
\bigl(r_d(t)-q_d\bigr)
-
\bigl(r_b(t)-q_b\bigr).
\end{aligned}
\label{eq:eps}
\end{equation}
\end{definition}

The normalization does not change the sign of either strength difference, so
\(\chi_{io}\) has the same sign as the derivative in \eqref{eq:assortderiv}.
Definition~\ref{def:contrast} gives the exact decompositions
\begin{equation}
\ell_a(t)-\ell_c(t)
=
\chi_{\mathrm{in}}+\varepsilon_{\mathrm{in}}(t),
\qquad
r_d(t)-r_b(t)
=
\chi_{\mathrm{out}}+\varepsilon_{\mathrm{out}}(t).
\label{eq:contrastdecomp}
\end{equation}
Each difference of Perron coordinates is a strength contrast plus a coordinate
error. Substituting these decompositions into the response of the Perron root gives
\begin{equation}
\rho'(t)
=
\delta\,
\frac{
\bigl(\chi_{\mathrm{in}}+\varepsilon_{\mathrm{in}}(t)\bigr)
\bigl(\chi_{\mathrm{out}}+\varepsilon_{\mathrm{out}}(t)\bigr)
}{
\eta(t)
},
\qquad
\eta(t)=\ell(t)^{\mathsf T}r(t)>0.
\label{eq:perronfactor}
\end{equation}
An error smaller in magnitude than its strength
contrast can neither reverse the sign of that contrast nor cancel it. When
both errors are smaller in magnitude than their strength contrasts, the Perron
root falls under a disassortative rewiring and rises under an assortative one,
because \(\delta\) and \(\eta(t)\) are positive.

\setcor{1a}
\begin{corollary}[Decline of the spectral depth]\label{cor:decline}
Suppose that Assumption~\ref{ass:pf} holds and that the rewiring is
disassortative, so that $\chi_{io}<0$.
\begin{enumerate}
\item[(i)]
If $\ell_a(t)-\ell_c(t)$ has the sign of $\chi_{\mathrm{in}}$ and $r_d(t)-r_b(t)$ has the sign of
$\chi_{\mathrm{out}}$ for every $t\in[0,1]$, then $\rho'(t)<0$ throughout the transfer and the
spectral depth $\mathcal{R}(W(t))$ decreases strictly.

\item[(ii)]
If there exist bounds
$\bar\varepsilon_{\mathrm{in}}$ and
$\bar\varepsilon_{\mathrm{out}}$ satisfying
\begin{equation}
\lvert\varepsilon_{\mathrm{in}}(t)\rvert
\le
\bar\varepsilon_{\mathrm{in}}
<
\lvert\chi_{\mathrm{in}}\rvert,
\qquad
\lvert\varepsilon_{\mathrm{out}}(t)\rvert
\le
\bar\varepsilon_{\mathrm{out}}
<
\lvert\chi_{\mathrm{out}}\rvert
\qquad
\text{for all }t\in[0,1],
\label{eq:uniform}
\end{equation}
then the hypothesis of part (i) holds, and
\begin{equation}
\mathcal{R}\bigl(W(1)\bigr)-\mathcal{R}\bigl(W(0)\bigr)
\le
-n\delta
\bigl(\lvert\chi_{\mathrm{in}}\rvert
      -\bar\varepsilon_{\mathrm{in}}\bigr)
\bigl(\lvert\chi_{\mathrm{out}}\rvert
      -\bar\varepsilon_{\mathrm{out}}\bigr).
\label{eq:declinebound}
\end{equation}
\end{enumerate}
Both parts hold under the symmetric transfer, with $2n\delta$ in place of
$n\delta$ in \eqref{eq:declinebound}.
\end{corollary}

\begin{proof}
Write $N(t)=(\ell_a(t)-\ell_c(t))(r_d(t)-r_b(t))$ and
$\eta(t)=\ell(t)^{\mathsf T}r(t)$, so that $\rho'(t)=\delta N(t)/\eta(t)$ by
Lemma~\ref{lem:perron}. For part (i), the rewiring is disassortative, so $\chi_{\mathrm{in}}$ and
$\chi_{\mathrm{out}}$ have opposite signs. Under the hypothesis of part (i), the two factors of
$N(t)$ have opposite signs as well. Hence $N(t)<0$, and $\rho'(t)<0$ because $\delta$ and $\eta(t)$ are positive.
By \eqref{eq:Rderivative}, the spectral depth $\mathcal{R}(W(t))$ decreases strictly throughout
the transfer.

For part (ii), by \eqref{eq:contrastdecomp}, the first factor of $N(t)$
is $\chi_{\mathrm{in}}+\varepsilon_{\mathrm{in}}(t)$. Since
$\lvert\varepsilon_{\mathrm{in}}(t)\rvert\le\bar\varepsilon_{\mathrm{in}}<\lvert\chi_{\mathrm{in}}\rvert$,
the first factor has the sign of $\chi_{\mathrm{in}}$ and magnitude at least
$\lvert\chi_{\mathrm{in}}\rvert-\bar\varepsilon_{\mathrm{in}}>0$. In the same way
the second factor, $\chi_{\mathrm{out}}+\varepsilon_{\mathrm{out}}(t)$, has the
sign of $\chi_{\mathrm{out}}$ and magnitude at least
$\lvert\chi_{\mathrm{out}}\rvert-\bar\varepsilon_{\mathrm{out}}>0$. So the hypothesis of part (i)
holds. The rewiring
is disassortative, so $\chi_{\mathrm{in}}$ and $\chi_{\mathrm{out}}$ have
opposite signs, and so do the two factors. Hence
\[
N(t)
\le
-
\bigl(\lvert\chi_{\mathrm{in}}\rvert
      -\bar\varepsilon_{\mathrm{in}}\bigr)
\bigl(\lvert\chi_{\mathrm{out}}\rvert
      -\bar\varepsilon_{\mathrm{out}}\bigr)
<0.
\]
The Perron vectors are positive unit vectors, so the Cauchy--Schwarz inequality
gives $0<\eta(t)\le1$, and $N(t)/\eta(t)\le N(t)$ because $N(t)<0$. It follows
that
\[
\rho'(t)
\le
-\delta
\bigl(\lvert\chi_{\mathrm{in}}\rvert
      -\bar\varepsilon_{\mathrm{in}}\bigr)
\bigl(\lvert\chi_{\mathrm{out}}\rvert
      -\bar\varepsilon_{\mathrm{out}}\bigr)
<0
\]
for every $t\in[0,1]$. By \eqref{eq:Rderivative}, integrating
the last estimate over $[0,1]$ and multiplying by $n$ gives
\eqref{eq:declinebound}.

Under the symmetric transfer, $\ell(t)=r(t)$ and $\eta(t)=1$, and Lemma~\ref{lem:perron} gives
$\rho'(t)=2\delta N(t)$. So the same argument applies with $2\delta$ in place of $\delta$.
\end{proof}

Part (i) of the corollary says that a disassortative rewiring lowers
the spectral depth whenever the left Perron vector
orders the two sources as their in-strengths do and the right Perron vector orders the two targets
as their out-strengths do. Part (ii) gives a condition under which the Perron vectors order the four rewired vertices in this
way,
together with a bound on the decline. The condition \eqref{eq:uniform} is easier to meet when the rewired
vertices differ more in strength and when the Perron vectors lie closer to the
normalized strength vectors. The transferred weight $\delta$ scales the
decline in \eqref{eq:declinebound}. But the transferred weight does not enter the strength contrasts.
And it enters the coordinate errors only through the Perron vectors along the
path.

Part (ii) holds with any valid bounds on the coordinate errors. Only the four rewired
vertices enter these errors, and the Cauchy--Schwarz inequality gives the bounds
\begin{equation}
\lvert\varepsilon_{\mathrm{in}}(t)\rvert
\leq
\sqrt{2}\,\lVert\ell(t)-p\rVert_2,
\qquad
\lvert\varepsilon_{\mathrm{out}}(t)\rvert
\leq
\sqrt{2}\,\lVert r(t)-q\rVert_2.
\label{eq:numall}
\end{equation}
In earlier work we derived Stewart--Sun perturbation bounds for the angle
between the normalized in-strength vector and the left Perron vector, and for the angle between
the normalized out-strength vector and the right Perron vector, for a class of
networks that includes weighted directed networks \cite{pv}. The strength and Perron vectors here are
unit vectors with the same positive orientation. This means that an angular bound also
controls their Euclidean distance through the identity
$\lVert u-v\rVert_2=2\sin(\angle(u,v)/2)$ for unit vectors $u$ and $v$. The Stewart--Sun bounds combine with \eqref{eq:numall} to control both coordinate errors when their hypotheses
hold uniformly along the transfer path.

But these global
bounds are conservative, for the sign of $\rho'(t)$ depends only on the coordinates at the four
rewired vertices, whereas each bound limits how far the whole Perron vector lies from the whole strength vector. The
global bounds also make the condition
\eqref{eq:uniform} hard to meet on a large network. The source contrast
$\lvert\chi_{\mathrm{in}}\rvert$ is at most the larger of $p_a$ and $p_c$, which is small unless one
of the two sources is a hub, whereas $\sqrt{2}\lVert\ell(t)-p\rVert_2$ is a distance between unit
vectors that need not shrink as the network grows. The same comparison holds at the targets,
between $\lvert\chi_{\mathrm{out}}\rvert$ and $\sqrt{2}\lVert r(t)-q\rVert_2$. Direct entrywise
estimates at the four rewired vertices, where available, give sharper bounds
on the coordinate errors.\footnote{In our computations none of the sampled rewirings meets the
norm-based version of the condition, whereas the signs of the differences of
Perron coordinates agree with the signs of the strength contrasts for most of the sampled rewirings at the start of the
transfer (Appendix~\ref{app:numerics}).}

\subsection{Bounds on the gaps}

A fall in the spectral depth of a real spectrum can leave some gaps wider than before. But the fall cannot raise the tail bound \eqref{eq:tail} at any threshold $u$, since that bound is nondecreasing in the spectral
depth. The fall also constrains every single gap, for every gap enters the spectral depth with a positive weight. No gap can exceed the spectral depth
divided by its weight. And in a real ordered spectrum the spectral depth beneath any eigenvalue
bounds the gaps below it from above in the same way. For a real ordered spectrum and $k\in\{1,\ldots,n-1\}$,
the spectral depth beneath $\lambda_k(t)$ is
\begin{equation}
\mathcal{R}_k\bigl(W(t)\bigr)
=
\sum_{r=k}^{n}
\bigl(\lambda_k(t)-\lambda_r(t)\bigr).
\label{eq:Rk}
\end{equation}

\setcor{1b}
\begin{corollary}[Depth beneath an arbitrary eigenvalue]\label{cor:anchor}
Suppose that the spectrum of $W(t)$ is real and ordered, and fix $k\in\{1,\ldots,n-1\}$. Then
\begin{equation}
\mathcal{R}_k\bigl(W(t)\bigr)
=
\sum_{i=k}^{n-1}(n-i)g_i(t),
\qquad
(n-k)g_k(t)
\le
\mathcal{R}_k\bigl(W(t)\bigr).
\label{eq:anchor}
\end{equation}
If the spectrum is real and ordered at the two endpoints $t=0$ and $t=1$, then,
with $\Delta\lambda_r=\lambda_r(1)-\lambda_r(0)$,
\begin{equation}
\mathcal{R}_k\bigl(W(1)\bigr)-\mathcal{R}_k\bigl(W(0)\bigr)
=
\sum_{r=1}^{k-1}\Delta\lambda_r
+
(n-k+1)\Delta\lambda_k .
\label{eq:exclusion}
\end{equation}
\end{corollary}

\begin{proof}
Because the spectrum is real and ordered,
$\lambda_k-\lambda_r=g_k+g_{k+1}+\cdots+g_{r-1}$ for every $r>k$. When these separations are summed
over $r=k,\ldots,n$, the gap $g_i$ appears $n-i$ times, once for each of the eigenvalues
$\lambda_{i+1},\ldots,\lambda_n$. The identity in \eqref{eq:anchor} follows. Every gap is
nonnegative and every coefficient $n-i$ is positive. The first term of the sum is $(n-k)g_k$, which
gives the inequality.

For the net change from $W(0)$ to $W(1)$, write
$\mathcal{R}_k=(n-k+1)\lambda_k-\sum_{r=k}^{n}\lambda_r$. Subtracting the values
at the two endpoints gives
\[
\mathcal{R}_k\bigl(W(1)\bigr)-\mathcal{R}_k\bigl(W(0)\bigr)
=
(n-k+1)\Delta\lambda_k-\sum_{r=k}^{n}\Delta\lambda_r .
\]
Trace preservation gives $\sum_{r=1}^{n}\Delta\lambda_r=0$, so
$-\sum_{r=k}^{n}\Delta\lambda_r=\sum_{r=1}^{k-1}\Delta\lambda_r$, which gives
\eqref{eq:exclusion}.
\end{proof}

Corollary~\ref{cor:anchor} bounds the gap $g_k$ through the spectral depth
beneath $\lambda_k$, with $g_k\le\mathcal{R}_k/(n-k)$. The bound is attained
exactly when every gap beneath $g_k$ vanishes, because
$\mathcal{R}_k=(n-k)g_k+\sum_{i=k+1}^{n-1}(n-i)g_i$ and every coefficient in
the last sum is positive. Because $\mathcal{R}_k\le\mathcal{R}$ whenever the spectrum is real and
ordered, every gap also satisfies
\[
g_k
\le
\frac{\mathcal{R}_k}{n-k}
\le
\frac{\mathcal{R}}{n-k}.
\]
Suppose that the hypotheses of Corollary~\ref{cor:decline} hold and that the spectrum stays real and
ordered, as it does under the symmetric transfer. Then the upper bound $\mathcal{R}/(n-k)$ on every
gap declines.

The identity \eqref{eq:exclusion} rests on the fixed trace. By that identity, the net change in the weighted sum $\mathcal{R}_k$ of the gaps beneath $\lambda_k$
from $W(0)$ to $W(1)$ depends only on the changes $\Delta\lambda_1,\ldots,\Delta\lambda_k$ of the
first $k$ eigenvalues. A decline in the Perron root forces $\mathcal{R}$ to
decline, since $\Delta\mathcal{R}=n\Delta\rho$. But the change in $\mathcal{R}_k$ for $k>1$ also
depends on $\Delta\lambda_2,\ldots,\Delta\lambda_k$, and these changes can offset the decline of the Perron root. Since the final depth $\mathcal{R}_k(W(1))$ is a weighted sum of nonnegative gaps, the net change
cannot be more negative than the initial depth $\mathcal{R}_k(W(0))$.

The upper bounds of Corollary~\ref{cor:anchor} are absolute, and on an absolute scale the same change may be
negligible for a large gap and substantial for a small one.
We can instead measure the contraction of each gap relative to its own
initial size, and weight it by the gap's share of the initial spectral depth.
The weighted contractions add up to at least the proportional fall in the
spectral depth. When the spectral depth falls by a proportion $\mathcal{D}$, the gaps
that each contract by more than a smaller proportion $\theta$ together carry at least a share of
the initial depth that depends only on $\mathcal{D}$ and $\theta$.

For two matrices $W(0)$ and $W(1)$ with real ordered spectra and $\mathcal{R}(W(0))>0$, write $I_+$
for the set of indices at which the gap of $W(0)$ is positive. For $i\in I_+$, the proportional
contraction $\phi_i$ of the $i$-th gap and its share $\psi_i$ of the initial depth are
\begin{equation}
\phi_i
=
\frac{g_i(0)-g_i(1)}{g_i(0)},
\qquad
\psi_i
=
\frac{(n-i)\,g_i(0)}{\mathcal{R}(W(0))}.
\label{eq:depthshare}
\end{equation}
We write $\mathcal{D}=1-\mathcal{R}(W(1))/\mathcal{R}(W(0))$ for the proportional fall in the
spectral depth and $Z=\sum_{i\notin I_+}(n-i)g_i(1)$ for the part of the final depth carried by
gaps that begin at zero.

\setcor{1c}
\begin{corollary}[Proportional gap contractions]\label{cor:proportional}
Suppose that $W(0)$ and $W(1)$ are the two ends of a rewiring or of a sequence of
rewirings, with spectra real and ordered, and that $\mathcal{R}(W(0))>0$. Then the weights $\psi_i$
sum to one over $I_+$, and for every $0\le\theta<1$,
\begin{equation}
\sum_{\substack{i\in I_+\\ \phi_i>\theta}}\psi_i
\ge
\frac{
\bigl(
\mathcal{D}+Z/\mathcal{R}(W(0))-\theta
\bigr)_+
}{
1-\theta
}
\ge
\frac{(\mathcal{D}-\theta)_+}{1-\theta}.
\label{eq:propshare}
\end{equation}
\end{corollary}

\begin{proof}
By Proposition~\ref{prop:global}, the spectral depth at each end is the weighted sum $\sum_{i=1}^{n-1}(n-i)g_i$ of the gaps. A gap that begins at zero contributes nothing to the initial depth, so the weights
$\psi_i$ sum to one over $I_+$, and $I_+$ is nonempty because that depth is positive.

In the weighted sum $\sum_{i\in I_+}\psi_i\phi_i$ the factor $g_i(0)$ cancels against the
denominator of $\phi_i$, leaving
\begin{equation}
\sum_{i\in I_+}\psi_i\phi_i
=
\frac{1}{\mathcal{R}(W(0))}
\sum_{i\in I_+}
(n-i)\bigl(g_i(0)-g_i(1)\bigr)
=
\mathcal{D}
+
\frac{Z}{\mathcal{R}(W(0))},
\label{eq:propbalance}
\end{equation}
the second equality because the gaps outside $I_+$ contribute nothing to the initial depth and
$Z$ to the final one. The quantity $Z$ is a sum of nonnegative terms.

For \eqref{eq:propshare} write $p_\theta$ for the left-hand side. Every gap at the end of the
transfer is nonnegative, so $\phi_i\le1$. On the complement of the set summed over,
$\phi_i\le\theta$. Splitting the weighted average \eqref{eq:propbalance} between the two sets
gives
\[
\mathcal{D}
+
\frac{Z}{\mathcal{R}(W(0))}
\le
p_\theta+(1-p_\theta)\theta.
\]
Rearranging for $p_\theta$, which is itself nonnegative, gives the first bound. Dropping $Z$ gives
the second bound.
\end{proof}

When the spectral depth falls, taking $\theta=0$ in \eqref{eq:propshare} shows that the gaps that
contract carry at least the share $\mathcal{D}$ of the initial depth. A higher threshold lowers the
bound, which reaches zero at $\theta=\mathcal{D}$.

\section{Assortativity and individual gaps}\label{sec:individual}

Proposition~\ref{prop:global} determines how the gaps taken together move and leaves open the question of whether a particular gap contracts or expands. For an admissible rewiring, the choice of the four rewired vertices decides whether the rewiring is assortative or disassortative,
and whether an individual gap contracts or expands depends on the weighted counts of walks between
those vertices.

\subsection{Eigenvalue velocities and counts of walks}

We follow how the rewiring moves the spectrum, as a whole and eigenvalue by eigenvalue, through the velocities of the eigenvalues along the
transfer path. Suppose that
the eigenvalues of \(W(t)\) are distinct, and write \(s_r(t)=\lambda_r'(t)\) for
\(r=1,\dots,n\). Lemma~\ref{lem:response} collects these velocities into a matrix element of the
resolvent of \(W(t)\), the scalar function
\begin{equation}
\mathcal{G}(z;t)
=
\delta\,y^{\mathsf T}\bigl(zI-W(t)\bigr)^{-1}x.
\label{eq:G}
\end{equation}
Its partial-fraction
representation is
\[
\mathcal{G}(z;t)
=
\sum_{r=1}^{n}
\frac{s_r(t)}{z-\lambda_r(t)}.
\]
The eigenvalues are the possible poles of \(\mathcal{G}\), and the residue at
a simple eigenvalue is the velocity of that eigenvalue. For \(\lvert z\rvert>\lVert W(t)\rVert_2\) the same function
has the expansion
\[
\mathcal{G}(z;t)
=
\sum_{q\geq0}\frac{\mu_q(t)}{z^{q+1}},
\qquad
\mu_q(t)
=
\delta\,y^{\mathsf T}W(t)^q x
=
\sum_{r=1}^{n}\lambda_r(t)^q s_r(t).
\]
The moment \(\mu_q(t)\) is
\(\delta\) times a signed count of weighted walks of length \(q\) from the
targets \(d\) and \(b\) to the sources \(a\) and \(c\), since
\(y^{\mathsf T}W^qx=(W^q)_{da}-(W^q)_{dc}-(W^q)_{ba}+(W^q)_{bc}\). The residues isolate the
velocities of individual eigenvalues. The moments, which need only powers of the weight matrix, give
weighted sums of the same velocities. Also note that the zeroth moment vanishes just as the fixed trace requires. The
four rewired vertices are distinct, so \(y^{\mathsf T}x=0\) and
\(\sum_{r=1}^{n}s_r(t)=\mu_0(t)=0\). Any movement of one part of the spectrum is offset by movement elsewhere.

The response of the Perron root in Lemma~\ref{lem:perron} is the residue of
\(\mathcal{G}\) at the Perron root. When the eigenvalues are distinct, Lemma~\ref{lem:recon} shows
that the first \(n\) moments \(\mu=(\mu_0,\dots,\mu_{n-1})^{\mathsf T}\) determine the complete
velocity vector \(s(t)=(s_1(t),\dots,s_n(t))^{\mathsf T}\) through the inverse of the Vandermonde
matrix \(\mathsf{V}(\lambda)\) of the spectrum, \(s=\mathsf{V}(\lambda)^{-1}\mu\). The same lemma
gives the velocity of each eigenvalue as \(s_r(t)=\delta\,y^{\mathsf T}L_r(W(t))x\), where \(L_r\)
is the full Lagrange polynomial of the spectrum associated with \(\lambda_r\). This polynomial
equals one at \(\lambda_r\) and zero at every other eigenvalue, so \(L_r(W(t))\) is the spectral
projector \(P_r\) of \(\lambda_r\).

\subsection{An exact criterion for the direction of a gap}

The velocity of the gap between two consecutive real ordered eigenvalues is
$g_i'(t)=s_i(t)-s_{i+1}(t)$. The gap contracts when $g_i'<0$ and expands when
$g_i'>0$.\footnote{Ordered gaps need a real spectrum, and beneath its
Perron root the spectrum of a directed network is rarely real. An ordering of complex eigenvalues by
modulus or real part may change under an arbitrarily small perturbation, so for such a spectrum we
follow two simple eigenvalue branches $\lambda_i(t)$ and $\lambda_j(t)$ continuously along the
transfer path. Their separation $z_{ij}(t)=\lambda_i(t)-\lambda_j(t)$ has the squared modulus
$D_{ij}(t)=\lvert z_{ij}(t)\rvert^2$. As long as the branches remain simple, they are analytic in $t$
\cite{kato}, and the radial derivative
\begin{equation}
D_{ij}'(t)=2\,\Real\bigl[\overline{z_{ij}(t)}\bigl(s_i(t)-s_j(t)\bigr)\bigr]
\label{eq:radial}
\end{equation}
tells whether the pair contracts ($D_{ij}'(t)<0$) or expands ($D_{ij}'(t)>0$). The radial derivative
measures only the radial component of the relative velocity, so the two eigenvalues may move rapidly
in the complex plane without moving closer together. For a conjugate pair, $D_{ij}'$ measures the
closing or widening of the imaginary separation.} By the inverse Vandermonde system of Lemma~\ref{lem:recon}, $g_i'=(e_i-e_{i+1})^{\mathsf T}
\mathsf{V}(\lambda)^{-1}\mu$. So the velocity of the $i$-th gap is a weighted sum of the first $n$
moments, and its weights are the entries of the difference between the $i$-th and $(i+1)$-st rows of
the inverse Vandermonde matrix. The same row
difference lists the coefficients, in the monomial basis, of the polynomial $Q_i=L_i-L_{i+1}$.

\begin{proposition}[Direction of a gap]\label{prop:criterion}
Suppose that the spectrum of $W(t)$ is real and simple, with
$\lambda_1(t)>\cdots>\lambda_n(t)$. Then
\begin{equation}
g_i'(t)
=
\delta\,y^{\mathsf T}Q_i\bigl(W(t)\bigr)x
=
\trace\bigl(Q_i(W(t))\,W'\bigr),
\label{eq:criterion}
\end{equation}
so the $i$-th gap contracts at $t$ if and only if
$y^{\mathsf T}Q_i(W(t))x<0$. Under the symmetric transfer the gap velocity is
$2\delta\,y^{\mathsf T}Q_i(W(t))x$, with the same sign.
\end{proposition}

\begin{proof}
The polynomial $Q_i$ takes the value $1$ at $\lambda_i$, the value $-1$ at
$\lambda_{i+1}$ and the value $0$ at every other eigenvalue, so
$Q_i(W)=P_i-P_{i+1}$. Lemma~\ref{lem:recon} gives
$s_r=\delta\,y^{\mathsf T}L_r(W)x$, and the difference $s_i-s_{i+1}$ is the
first identity in \eqref{eq:criterion}. The second follows from cyclicity of
the trace after substituting $W'=\delta xy^{\mathsf T}$. When $W$ is symmetric so is
$Q_i(W)$, and the velocity $E=\delta(xy^{\mathsf T}+yx^{\mathsf T})$ of the symmetric transfer
gives $\trace(Q_i(W)E)=2\delta\,y^{\mathsf T}Q_i(W)x$.
\end{proof}

By Proposition~\ref{prop:criterion}, whether a rewiring contracts a given gap is decided by one
number, a weighted sum of the counts of walks from the two targets to the two sources. For a rewiring that meets the hypothesis of Corollary~\ref{cor:decline}(i), the same number tells us
whether the perturbation that lowers the spectral depth of a real spectrum also contracts the given
gap.

The criterion is exact and requires no eigenvector calculation. It gives the direction at one point
of the path, and it is costly for a large matrix. The polynomial $Q_i$ depends on the complete
spectrum and uses the moments up to order $n-1$ in the monomial basis. Lemma~\ref{lem:nonid} shows that this
order cannot be lowered, since the moments up to order $n-2$ remain compatible both with the contraction and with the
expansion of any given gap.

A criterion that uses fewer moments needs some structure in the spectrum, such
as a few of the largest eigenvalues being sufficiently separated from the rest of the spectrum. We call these eigenvalues the retained eigenvalues. The velocity of an
eigenvalue $\lambda_r$ is $s_r=\trace(P_rW')$, where $P_r$ is its spectral projector. To approximate
the velocity of a retained eigenvalue without the rest of the spectrum, we replace $P_r$ by $f(W)$
for a scalar function $f$ that is close to one at $\lambda_r$ and close to zero at every other
eigenvalue. We call such a function a filter. Appendix~\ref{app:lemmas} defines the two filters we
use. The polynomial filter \eqref{eq:cheb} needs a real spectrum whose retained eigenvalues lie
above an interval that contains the other eigenvalues. The rational filter \eqref{eq:rational}
needs only a circle around the retained eigenvalue that excludes the other eigenvalues.
Lemma~\ref{lem:filter} bounds their errors in \eqref{eq:filtbound} and \eqref{eq:ratbound}. Both
bounds are proportional to the condition number of an eigenbasis, which can equal one only when $W$
is normal.\footnote{In our computations the rational filter recovers the velocities of all the
retained eigenvalues to within its error bound. In floating-point arithmetic, the polynomial filter
recovers only the velocity of the Perron root (Appendix~\ref{app:numerics}).}

The error bounds \eqref{eq:filtbound} and \eqref{eq:ratbound} place each
approximate velocity $\widehat s_r$ within a known distance of the exact velocity
$s_r$. So the error in the approximate gap velocity $\widehat s_i-\widehat s_{i+1}$ is at most the sum of the two distances. The approximate gap velocity has the sign of $g_i'$
whenever its magnitude exceeds that sum.

\subsection{Patterns of contraction and expansion}
\label{sec:patterns}

The exact criterion of Proposition~\ref{prop:criterion} for the direction of a gap uses every moment
$\mu_q$ up to order $n-1$. But the moments up to order $M$, combined with a falling Perron root,
still constrain the gaps. What is intriguing is that the low-order moments force a subsequence of gaps to alternate between
contraction and expansion down the spectrum, even though they do not determine every eigenvalue
velocity.

\begin{proposition}[Alternating contraction and expansion]\label{prop:patterns}
Suppose that Assumption~\ref{ass:pf} holds, that the spectrum at a fixed $t_0\in[0,1]$ is real and
simple with $\lambda_1(t_0)>\lambda_2(t_0)>\cdots>\lambda_n(t_0)$, and that $\rho'(t_0)<0$.
Write $s_r=\lambda_r'(t_0)$ and $\mu_q=\delta\,y^{\mathsf T}W(t_0)^qx$, fix
$M\in\{1,\ldots,n-2\}$, and suppose that
\begin{equation}
\mu_q=0
\qquad
(1\le q<M),
\qquad
\mu_M\ge0.
\label{eq:patternhyp}
\end{equation}
When $M=1$ the vanishing conditions are empty. Then there exist indices
$1\le i_1<i_2<\cdots<i_{M+1}\le n-1$ such that
\begin{equation}
g_{i_j}'(t_0)<0
\quad\text{when $j$ is odd},
\qquad
g_{i_j}'(t_0)>0
\quad\text{when $j$ is even}.
\label{eq:patternsigns}
\end{equation}
The same holds under the symmetric
transfer.
\end{proposition}

\begin{proof}
Suppress $t_0$ from the notation. By Lemma~\ref{lem:response},
$\mu_q=\sum_{r=1}^{n}\lambda_r^qs_r$ for every $q\ge0$, and trace preservation
gives $\mu_0=\sum_{r=1}^{n}s_r=0$. Under the symmetric transfer,
Lemma~\ref{lem:response} gives $\sum_{r=1}^{n}\lambda_r^qs_r=2\mu_q$ instead, which changes neither
the hypotheses \eqref{eq:patternhyp} nor the argument that follows.

Let $k$ be the number of sign changes in the sequence of velocities $s_1,\ldots,s_n$, taken in order
with its zero entries removed. The first entry of this sequence is $s_1=\rho'<0$. We first show
that $k\ge M+1$, and we do so by contradiction. Suppose instead that $k\le M$. At each sign change,
choose a number $\xi_\ell$ strictly between the eigenvalues associated with the two consecutive
nonzero velocities across that sign change. Define $p(z)=-\prod_{\ell=1}^{k}(z-\xi_\ell)$, where the
empty product is $1$. Because every $\xi_\ell<\lambda_1$, we have $p(\lambda_1)<0$. As $r$
increases, the sign of $p(\lambda_r)$ changes exactly where the sign of the nonzero velocity $s_r$
changes, and $s_1<0$, so $p(\lambda_r)s_r>0$ whenever $s_r\ne0$. Hence
\[
\sum_{r=1}^{n}p(\lambda_r)s_r>0.
\]
But the moment conditions force the same sum to be nonpositive. Since
$\sum_r\lambda_r^qs_r=\mu_q$, the sum is the combination of the moments with the coefficients of $p$
as weights, and $\mu_q=0$ for every $q<M$. If $k<M$, then $p$ has degree at most $M-1$, so the sum
is zero. If $k=M$, then $p(z)=-z^M$ plus terms of lower degree, so the sum equals $-\mu_M\le0$. Both
cases contradict the positivity of the sum, so $k\ge M+1$.

We now use the sign changes to find the indices of the proposition. Since $k\ge M+1$, the nonzero
velocities form at least $M+2$ successive blocks of constant sign. Choose indices
$r_0=1<r_1<\cdots<r_{M+1}$, one from each of the first $M+2$ blocks. Since $s_{r_0}<0$, the signs
of $s_{r_0},\ldots,s_{r_{M+1}}$ alternate, with $s_{r_j}<0$ for even $j$ and $s_{r_j}>0$ for odd
$j$. For each $j$ the gap velocities telescope,
\[
\sum_{i=r_{j-1}}^{r_j-1}g_i'
=
s_{r_{j-1}}-s_{r_j}.
\]
The sum is negative when $j$ is odd and positive when $j$ is even. So each of the $M+1$ disjoint
intervals of indices $\{r_{j-1},\ldots,r_j-1\}$ contains at least one gap derivative with the sign of
its sum, and selecting one such index from each interval gives $i_1<i_2<\cdots<i_{M+1}$.
\end{proof}

With $M=1$, the proposition says that when the Perron root falls and
$\mu_1\ge0$, a contracting gap lies above an expanding gap somewhere in the
ordered spectrum. The first moment,
$\mu_1=\delta\,y^{\mathsf T}Wx=\delta(w_{da}-w_{dc}-w_{ba}+w_{bc})$, depends only
on the four entries of $W$ that join the two targets $d$ and $b$ to the two
sources $a$ and $c$. The edges $d\to a$ and $b\to c$ run back along the two crossed edges, and the
edges $b\to a$ and $d\to c$ run back along the two donor edges. The condition $\mu_1\ge0$ asks the
edges $d\to a$ and $b\to c$ to carry at least as much weight as the edges $b\to a$ and $d\to c$. In a symmetric network an edge and the edge that runs
back along it carry the same weight, so $\mu_1\ge0$ says that the crossed edges
already carry at least as much weight as the donor edges.

For larger $M$, the vanishing of further moments forces longer alternating
sequences. At $M=n-2$ the $M+1$ selected indices are all $n-1$ gaps, so every odd-numbered gap
contracts and every even-numbered gap expands. The moment conditions are restrictions in their
own right, for neither
disassortativity nor a falling Perron root implies them. And the falling Perron root $s_1=\rho'<0$
is the hypothesis that forces the alternation. By Lemma~\ref{lem:nonid}, the low-order moments alone cannot settle the direction of any gap, since
they are compatible both with its contraction and with its expansion.

In all the trajectories of our computations the crossed edges carry no weight at the start of the
rewiring. On a symmetric network such a rewiring has $\mu_1<0$ at $t=0$, so the proposition does not
apply at the start of the rewiring for any $M$. On a sparse directed network the four edges that run
back from the targets to the sources are usually absent, so that $\mu_1=0$, but its spectrum is
rarely real.

\subsection{Contraction throughout a completed transfer}\label{sec:mesh}

Proposition~\ref{prop:criterion} and the radial derivative \eqref{eq:radial} give the direction in
which a gap or a pair of eigenvalues moves at a single point of the path. But a rewiring transfers
a finite weight. What remains to be shown is when a gap or a pair contracts at every point of the
path, and so over the whole transfer from the original matrix to the rewired matrix. The sign of a
velocity at one point of the path need not persist along the whole path, because the velocities $g_i'$ and $D_{ij}'$ change as the
eigenvectors rotate along the path. Lemma~\ref{lem:curv} provides curvature constants
$C_i$ and $F_{ij}$ such that $\lvert g_i''(t)\rvert\le C_i$ and
$\lvert D_{ij}''(t)\rvert\le F_{ij}$. With these bounds we settle the sign of a velocity on the
whole path from its values, or upper bounds on them, at the $J+1$ equally spaced points $t_m=m/J$ of
the path. Here $J$ is a positive integer of our choosing. The bound on the second derivative then
controls the velocity between these points.

\begin{proposition}[Finite-mesh condition for contraction]\label{prop:certified}
Let $J$ be a positive integer, $h=1/J$ and $t_m=mh$ for $0\le m\le J$.
\begin{enumerate}
\item[(i)]
Suppose that the branches $\lambda_1(t)>\cdots>\lambda_{k+1}(t)$ are real and simple on
$[0,1]$, that $C_i$ is the curvature constant of Lemma~\ref{lem:curv} for
$\lambda_i$ and $\lambda_{i+1}$, and that $g_i'(t_m)\le\overline v_{i,m}$ at every mesh point, for
$i\le k$. If
\begin{equation}
\max_{0\le m\le J}\overline v_{i,m}
<
-\frac{C_i h}{2},
\qquad
i=1,\ldots,k,
\label{eq:mesh}
\end{equation}
then $g_i'(t)<0$ for every $t\in[0,1]$ and every $i\le k$.

\item[(ii)]
Suppose that each pair of branches $\lambda_i,\lambda_j$ in a finite collection is
distinct and simple on $[0,1]$, that $F_{ij}$ is the curvature constant
of Lemma~\ref{lem:curv} for the pair, and that
$D_{ij}'(t_m)\le\overline d_{ij,m}$ at every mesh point. If
$\max_{0\le m\le J}\overline d_{ij,m}<-F_{ij}h/2$ for every pair in the
collection, then every pair contracts strictly throughout $[0,1]$.
\end{enumerate}
\end{proposition}

\begin{proof}
For part (i), fix $i\le k$. Since $\lambda_i$ and $\lambda_{i+1}$ are simple on
$[0,1]$, the gap velocity $g_i'$ is continuously differentiable, and
Lemma~\ref{lem:curv} gives $\lvert g_i''(t)\rvert\le C_i$. Every $t\in[0,1]$
lies within $h/2$ of some mesh point $t_m$, and the mean value theorem gives
\[
g_i'(t)
\le
g_i'(t_m)+C_i\lvert t-t_m\rvert
\le
\overline v_{i,m}+\frac{C_i h}{2}
\le
\max_{0\le m'\le J}\overline v_{i,m'}+\frac{C_i h}{2}
<0.
\]
Each of the first $k$ gaps contracts strictly throughout the transfer.

For part (ii), fix one pair in the collection. Lemma~\ref{lem:curv} gives
$\lvert D_{ij}''(t)\rvert\le F_{ij}$, and the same argument applied to
$D_{ij}'$ gives
$D_{ij}'(t)\le\max_{0\le m\le J}\overline d_{ij,m}+F_{ij}h/2<0$. The pair
contracts strictly throughout the transfer, and the argument applies to every
pair in the collection.
\end{proof}

Put simply, a gap that contracts faster
than the curvature allowance $C_ih/2$ at every mesh point contracts throughout
the transfer. The finite-mesh condition is sufficient but not necessary, so a gap that fails it
may still contract at every point of the transfer. The condition is hardest to meet when the two
eigenvalues that bound the gap lie close to the rest of the spectrum or have large condition
numbers $\kappa_i$ and $\kappa_{i+1}$, for the curvature constant $C_i$ of Lemma~\ref{lem:curv}
grows with the norms of their reduced resolvents $S_i$ and $S_{i+1}$ and with $\kappa_i$ and
$\kappa_{i+1}$. The norm of the reduced resolvent
grows as an eigenvalue
approaches the rest of the spectrum, and the condition number $\kappa_r$ grows
as the left and right eigenvectors of $\lambda_r$ approach orthogonality. For a normal matrix such
as a symmetric weight matrix, $\kappa_r=1$ and $\lVert S_r\rVert_2=1/\gamma_r$, where $\gamma_r$ is
the distance from $\lambda_r$ to the rest of the spectrum. For a directed nonnormal matrix both
quantities can be far larger.

The initial
gap velocity $g_i'(0)$ is proportional to $\delta$, whereas $C_i$ is $8\delta^2$ times
a sum of two suprema taken along a path that itself shortens as $\delta$ falls, so $C_i$ shrinks
at least as fast as $\delta^2$. A smaller transfer shrinks the
curvature allowance faster than the gap velocity, and a gap whose velocity is negative at the
start of the path meets the finite-mesh condition once the transfer is small
enough.\footnote{Even without a mesh, if $g_i'(0)<-C_i/2$, then $g_i(1)-g_i(0)\le g_i'(0)+C_i/2<0$
by the curvature bound. So the gap is smaller at the end of the transfer than at its start, although
its velocity may change sign along the way.}

Requiring the first $k$ gaps to contract together is stronger than requiring
one gap to contract. Since $g_i'=s_i-s_{i+1}$, the first $k$ gaps contract
together only if $s_1<s_2<\cdots<s_{k+1}$ throughout the transfer. The
eigenvalue velocities are the residues of the same scalar function
\eqref{eq:G} and cannot be chosen independently.

The finite-mesh condition \eqref{eq:mesh} bounds the gap velocity away from
zero, and the contractions it implies add up across successive transfers.

\setcor{4}
\begin{corollary}[Contraction over a sequence of rewirings]\label{cor:repeat}
Suppose that each $W^{(\tau+1)}$ is obtained from $W^{(\tau)}$ by a directed or a symmetric
transfer,
\[
W^{(\tau+1)}=W^{(\tau)}+\delta^{(\tau)}B^{(\tau)}
\quad\text{or}\quad
W^{(\tau+1)}=W^{(\tau)}+\delta^{(\tau)}\bigl(B^{(\tau)}+(B^{(\tau)})^{\mathsf T}\bigr),
\qquad \tau=0,\ldots,N-1.
\]
Suppose also that every transfer satisfies the hypotheses of
Proposition~\ref{prop:certified}(i) for the first $k$ gaps, with mesh width
$h^{(\tau)}$. Put $\zeta_{i,\tau}=-\max_m\overline v_{i,m}^{(\tau)}-C_i^{(\tau)}h^{(\tau)}/2$, which is
positive by \eqref{eq:mesh}. Then
\begin{equation}
g_i\bigl(W^{(N)}\bigr)
\le
g_i\bigl(W^{(0)}\bigr)
-
\sum_{\tau=0}^{N-1}\zeta_{i,\tau},
\qquad
i=1,\ldots,k.
\label{eq:repeat}
\end{equation}
\end{corollary}

\begin{proof}
Fix $i\le k$ and a transfer $\tau$. The proof of Proposition~\ref{prop:certified}(i)
gives $g_i'(t)\le-\zeta_{i,\tau}$ throughout that transfer, and integrating over
$[0,1]$ gives $g_i(W^{(\tau+1)})-g_i(W^{(\tau)})\le-\zeta_{i,\tau}$. The endpoint of one
transfer is the starting point of the next, so summing over $\tau=0,\ldots,N-1$
makes the intermediate values cancel and gives \eqref{eq:repeat}.
\end{proof}

For a sequence of disassortative rewirings selected through the strength
contrasts of Definition~\ref{def:contrast}, Corollary~\ref{cor:repeat} adds up
the contraction established at each step into a lower bound on the total fall
in the gap over the whole sequence. Once a gap closes, its two eigenvalues are no longer simple, and the sequence to which Corollary~\ref{cor:repeat} applies ends at that rewiring.

\section{Computational experiments}\label{sec:computations}

We now use computational experiments to relax the assumptions we had to make to derive the
analytical results of Sections~\ref{sec:whole} and~\ref{sec:individual}. The most significant of
these is the assumption that the Perron vectors rank the four rewired vertices as their strengths would rank them. Since our analytical results say little
about the exact magnitude of the changes in the spectral gaps in response to assortativity, we also
use the experiments to measure the size of the effects.

The networks we use for the computational experiments come from the configuration model, which
draws a degree for each vertex and pairs the ends of the edges at random. Their degrees follow a
truncated power law, except in the homogeneous controls, whose degrees follow a Poisson
distribution. Their edge weights are lognormal. Each experiment runs on 100 independently drawn networks with between $600$ and
$20{,}000$ vertices, and Appendix~\ref{app:numerics} repeats the experiments on networks of up
to $320{,}000$ vertices.  We measure assortativity by the normalized coefficient
$r_{io}=C_{io}/(\sigma_{\mathrm{in}}\sigma_{\mathrm{out}})$, where $\sigma_{\mathrm{in}}$ and
$\sigma_{\mathrm{out}}$ are the edge-weighted standard deviations of source in-strength and target
out-strength.\footnote{Since a transfer preserves every strength, $\sigma_{\mathrm{in}}$ and
$\sigma_{\mathrm{out}}$ stay constant along it. Under Assumption~\ref{ass:pf} both are positive unless
all in-strengths or all out-strengths are equal, so $r_{io}$ moves in the same direction as
$C_{io}$.}

Each trajectory is a sequence of rewirings that transfer the largest admissible weight. On a
weighted network such a rewiring changes the number of edges even though it changes no strength.
With continuous weights the two donor edges differ in weight, and the transferred weight is the
weight of the lighter one. So the lighter donor edge disappears, the heavier donor edge keeps the
difference, and the two crossed edges, which were absent before, appear. Suppose, for instance, that
the edge from $a$ to $b$ carries less weight than the edge from $c$ to $d$. The rewiring removes the
edge from $a$ to $b$, lowers the weight of the edge from $c$ to $d$ by the weight removed, and
creates the edges from $a$ to $d$ and from $c$ to $b$. The network ends with one more edge than it
began with, while every vertex keeps its in-strength and its out-strength. We call the set of edges
that carry positive weight the support of the network, so each rewiring on a weighted network adds
one edge to the support. On a binary network the two donor edges carry the same weight, both
disappear, and the support keeps its size. 

Except in the experiments on homogeneous networks, we compare rewirings selected by the sign of
their effect on assortativity with unselected rewirings, accepted without regard to that sign, which
add edges at the same rate. In the experiments on the spectral depth, on the shape of the gap
distribution and on gaps of length zero, we also repeat the rewirings on binary networks. We call
each combination of a selection (disassortative, assortative or unselected) with a kind of network
(weighted or binary) a rule. Along each trajectory we record the spectrum, or on the largest
networks the Perron root and the trace, at checkpoints spaced geometrically in the number of
rewirings from the first rewirings to the last. Appendix~\ref{app:design} describes the networks,
the controls and the numerical conventions in detail.

In Section~\ref{sec:gapdist} we select rewirings by the sign of their effect on assortativity alone,
follow the spectral depth along long sequences of them and measure where in the spectrum the gaps
contract. In Section~\ref{sec:individualnum} we turn to the first gap, which governs how fast the network converges to its long-run state, and measure how often rewiring contracts it and by how much. In Section~\ref{sec:origin} we measure the share of
gaps of length zero, which repeated eigenvalues create. Since ordered gaps need a real spectrum, the experiments on the shape of the gap distribution, on
the first gap and on gaps of length zero use symmetric networks, whereas the experiment on the
spectral depth uses directed networks. In Section~\ref{sec:individualnum} we also follow pairs of complex eigenvalues on directed networks
in place of ordered gaps, with the details in Appendix~\ref{app:pairstest}.

\subsection{The spectral depth and the shape of the spectral gap distribution}\label{sec:gapdist}

\begin{figure}[h!]\centering
  \includegraphics[width=0.96\linewidth]{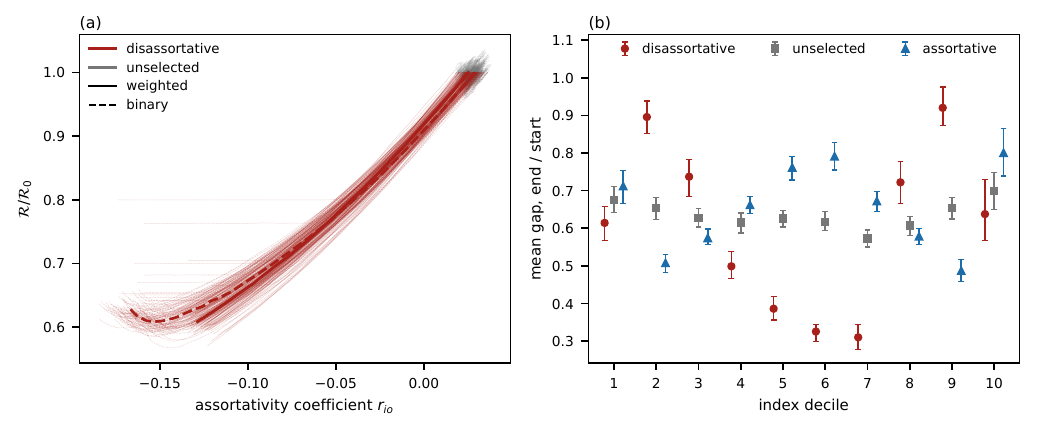}
  \caption{Spectral depth and gaps under rewiring. (a) The spectral depth $\mathcal{R}$ relative
  to its initial value against the assortativity coefficient $r_{io}$ on directed networks of
  $20{,}000$ vertices, with one thin line for each of 100 networks per rule and the median over
  the networks of each rule in bold. (b) The mean of the
  gaps $g_{11},\dots,g_{n-2}$ in each index decile after $30{,}000$ rewirings relative to its
  initial value on symmetric heavy-tailed networks of $1{,}200$ vertices, with decile 1 at the
  top of the spectrum. Markers give the mean over 100 networks, and bars give the range from the 5th to
  the 95th percentile.
  }
  \label{fig:gapdist}
\end{figure}

In this subsection we measure how much a change in assortativity moves the spectral depth, and where
in the spectrum the gaps contract.\footnote{The experiments impose no condition on the Perron
vectors. Corollary~\ref{cor:decline}(ii) ensures that the Perron vectors rank the four rewired
vertices as their strengths do when the strength contrasts exceed the bounds on the coordinate
errors, the amounts by which the differences of Perron coordinates depart from the strength
contrasts. The bounds we can compute are larger than the strength contrasts of the rewirings we
sample (Appendix~\ref{app:ranking}), so the experiments measure what disassortative rewiring does
to the spectral depth with the ranking left to take care of itself.} We run four experiments, each on 100 directed heavy-tailed networks of $20{,}000$ vertices and
each over about $200{,}000$ rewirings. The experiments combine disassortative or unselected
rewirings with weighted or binary networks. Figure~\ref{fig:gapdist}(a) presents the results. Since each rewiring moves $r_{io}$ by an exact
amount, each line shows how the spectral depth of one network responds to assortativity.

The experiments show that disassortative rewiring lowers the spectral depth by about two fifths, on
weighted and binary networks alike, while unselected rewiring leaves it nearly unchanged. Along the weighted disassortative trajectories, the mean coefficient $r_{io}$ falls from $0.026$ to $-0.128$.
The spectral depth falls by $39.0$ percent on average, and by between $36.7$ and $41.0$ percent on
nine networks in ten. In 99 of the 100 weighted trajectories the spectral depth reaches its minimum
at the final checkpoint. The binary disassortative trajectories lower the spectral depth by $36.7$
percent on average and follow nearly the same curve against $r_{io}$, with a late rebound of about
two percent of the initial depth. The unselected trajectories stay within $4.6$ percent of their
initial depth throughout, and within $3$ percent on 98 of the 100 weighted networks and on 82 of the
100 binary networks. By
Proposition~\ref{prop:global}, every fall in $\mathcal{R}$ is $n$ times a fall in the Perron root.\footnote{The growth of the support does not explain the decline. The unselected weighted trajectories grow
the support at the same rate as the disassortative trajectories and show no decline. And the binary
disassortative trajectories show the decline without growing the support at all.}

The fall of about two fifths persists as the networks grow, except where the diagonal stops it. More specifically, a
nonnegative matrix has a Perron root at least as large as its largest diagonal entry, so the
lognormal diagonal puts a floor under the Perron root, and the largest of $n$ lognormal draws grows
with $n$. Without a diagonal, the weighted disassortative trajectories lower the spectral depth by
$38.1$, $39.6$, $40.4$ and $40.4$ percent on networks of $5{,}000$, $20{,}000$, $80{,}000$ and
$320{,}000$ vertices, and the spread across networks narrows to between $39.9$ and $40.9$ percent at
the largest size. With the diagonal, the Perron root ends at this floor on 7 of the 10 weighted and
8 of the 10 binary networks of $320{,}000$ vertices, where the fall shrinks to $36.2$ and $29.4$ per
cent on average (Appendix~\ref{app:ranking}). The floor also accounts for the smallest falls among the
networks of $20{,}000$ vertices.

To see where in the spectrum the gaps contract, we run three experiments, each on 100 symmetric
heavy-tailed networks of $1{,}200$ vertices and each over $30{,}000$ rewirings. The experiments use disassortative, assortative and
unselected rewirings on weighted networks, and a binary control repeats them with every
off-diagonal weight equal to one. Disassortative rewiring moves the mean $r_{io}$ from $0.03$ to
$-0.36$ and lowers the spectral depth to $0.72$ of its initial value on average. By
Corollary~\ref{cor:proportional}, the gaps that each lose more than a tenth of their length under
disassortative rewiring carry a share of the initial depth whose lower bound averages about 20 per
cent across the 100 networks. Assortative rewiring raises $r_{io}$ to $0.39$ and $\mathcal{R}$ to
$1.43$ times its initial value.

Figure~\ref{fig:gapdist}(b) presents the results across the spectrum, with the gaps
$g_{11},\dots,g_{n-2}$ divided by index into ten deciles. The deciles leave out the last gap and the
ten leading gaps, which carry the largest weights in the spectral depth of a real spectrum. The experiments show that disassortative and assortative rewiring contract the gaps most strongly in
opposite parts of the spectrum. Unselected rewiring lowers the mean gap in every decile to between
$0.57$ and $0.70$ of its initial value. Because the omitted gaps also contribute to the spectral
depth, these decile ratios alone do not determine its change under unselected rewiring.
Disassortative rewiring contracts the gaps in the middle of the spectrum most and the gaps just inside
its two ends least. Under disassortative rewiring the fifth to seventh deciles fall to between $0.31$
and $0.39$, about half the unselected ratio. The second, third, eighth and ninth deciles stay between
$0.72$ and $0.92$, above the unselected ratio. The first and tenth deciles fall to roughly the
unselected level. Assortative rewiring produces the opposite profile. Under assortative rewiring the
strongest contraction is in the second and ninth deciles, where the ratios are about $0.50$, below
the unselected ratio. The middle deciles stay above the unselected ratio. Since on 30 networks of $2{,}400$ vertices and
20 of $4{,}800$, with the number of rewirings in proportion to the number of vertices, every decile
ratio of every rule lies within $0.05$ of its value at $1{,}200$ vertices
(Appendix~\ref{app:design}), the profile does not depend on the size of the network.

The binary control separates the selection from the growth of the support. On binary networks unselected rewiring leaves the mean gap of every decile at its initial value. The contraction of every decile under unselected weighted
rewiring comes from the growth of the support. Disassortative rewiring of binary networks
still contracts the middle of the spectrum most, with the fourth to seventh deciles falling to
between $0.24$ and $0.42$ of their initial values. But it widens the first, second, ninth and tenth
deciles, to between $1.64$ and $1.79$ times their initial values. Assortative rewiring of binary
networks contracts the second, ninth and tenth deciles most. So the selection sets where in the
spectrum the gaps contract most, and the growth of the support adds a contraction of every decile.

\subsection{The first spectral gap}\label{sec:individualnum}

We now turn from the gaps taken together to the first gap, which sets how fast the network
converges to its long-run state. We ask how often disassortative and assortative rewirings contract the first gap, and by how much
they change it, over one rewiring and over a long sequence of rewirings. We also ask how closely the
lower bound we derive for its fall tracks the realized fall along a sequence of transfers.

We run three experiments, each on 100 symmetric heavy-tailed networks of $600$ vertices, and we
also follow the first gap along the trajectories of Section~\ref{sec:gapdist}. The first experiment
enumerates every admissible rewiring of each network, about ten million per network, classifies each
by the sign of $\chi_{io}$ and records whether the first gap contracts at the start of the transfer,
which Proposition~\ref{prop:criterion} decides. Table~\ref{tab:firstgap} reports the mean share
across the 100 networks and the range from the 5th to the 95th percentile. The share under unselected rewiring is the share of all admissible rewirings that contract the first
gap. The second experiment draws $120$
admissible rewirings on each network, transfers the largest admissible weight, and records the
change of the first gap over the completed transfer. The third experiment applies $40$
transfers in sequence to each network, each at a thousandth of the largest admissible weight and
each meeting the finite-mesh condition for the first gap.

\begin{table}[h!]
\centering
\small
\caption{Share of admissible rewirings that contract the first gap at the start of the transfer, on
100 symmetric heavy-tailed networks of $600$ vertices, with every admissible rewiring of each network
counted. The range runs from the 5th to the 95th percentile across networks.}
\label{tab:firstgap}
\begin{tabular}{@{}lcc@{}}
\toprule
Rewiring & Mean share & Range across networks\\
\midrule
Disassortative & $0.605$ & $0.569$--$0.648$\\
Assortative & $0.260$ & $0.244$--$0.275$\\
Unselected & $0.435$ & $0.419$--$0.452$\\
\bottomrule
\end{tabular}
\end{table}

The experiments show that a disassortative rewiring contracts the first gap more than twice as
often as an assortative rewiring, and more often than an unselected rewiring
(Table~\ref{tab:firstgap}). This ordering of the three shares holds on every one of the 100 networks. Our analytical results explain both why the sign of a rewiring's effect on assortativity changes
how often the first gap contracts and why it leaves the direction of any one rewiring open. The first gap contracts exactly when
the Perron root has the smaller velocity of the two eigenvalues that bound it. A disassortative
rewiring that lowers the Perron root adds a negative term to the velocity of the first gap. The negative term
makes contraction more likely. On the networks of $600$ vertices, $87$ percent of the disassortative rewirings lower the Perron root, and $65$ percent of
these contract the first gap, whereas only $31$ percent of the disassortative rewirings that raise
the Perron root do so. Assortative rewirings raise the Perron root in $77$ percent of cases, and so
contract the first gap least often. But the velocity of the second eigenvalue rests on the counts of walks
of every length and on the whole spectrum, and the counts of all but the longest walks are
compatible with both the contraction and the expansion of the first gap (Lemma~\ref{lem:nonid}).
Table~\ref{tab:firstgap} shows this pattern. Disassortative and assortative rewiring move the share
of contracting rewirings in opposite directions from the unselected share, but neither share comes close to one or to zero.\footnote{On larger networks the term from the
Perron root weighs more. The share of disassortative rewirings that contract the first gap rises
from $0.61$ at $600$ vertices to $0.72$ at $9{,}600$ vertices, and the assortative share falls from
$0.26$ to $0.22$, though neither share approaches one or zero at these sizes
(Appendix~\ref{app:criterion}).}

\begin{figure}[h!]\centering
  \includegraphics[width=0.96\linewidth]{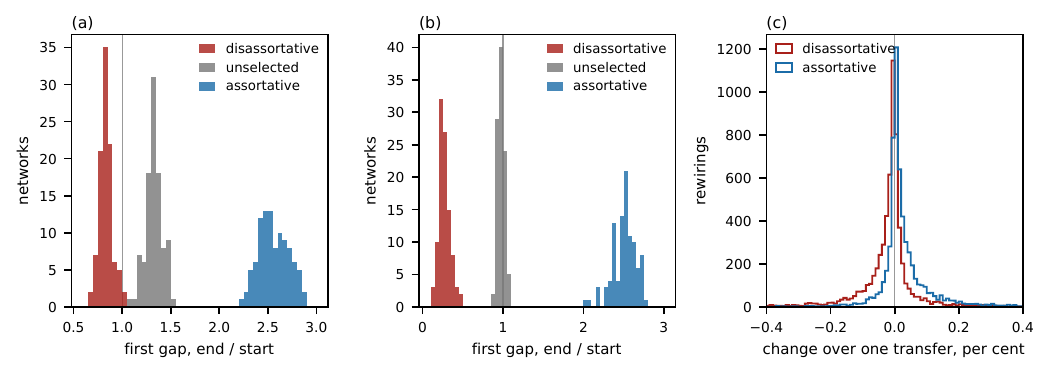}
  \caption{The size of the change in the first gap. (a) The first gap after $30{,}000$ rewirings
  relative to its initial value on the 100 symmetric heavy-tailed networks of $1{,}200$ vertices
  with lognormal weights of Section~\ref{sec:gapdist}, one count per network under each rule.
  (b) The same on the binary networks, whose support does not grow. (c) The change of the first
  gap over one completed transfer of the largest admissible weight, in percent of its initial
  value, over $120$ transfers on each of the 100 networks of $600$ vertices of
  Table~\ref{tab:firstgap}, for disassortative and for assortative rewirings. The range shown contains $98$ percent of the transfers of each sign.}
  \label{fig:firstgap}
\end{figure}

We now measure by how much the first gap changes. Figure~\ref{fig:firstgap}(c) presents the change
of the first gap over one completed transfer of the largest admissible weight, in percent of its
initial value, over the $120$ transfers on each of the 100 networks. The change of the first gap over one transfer is small, with a mean of $-0.024$ percent under disassortative rewiring and of $0.036$ percent under assortative
rewiring. Nine disassortative rewirings in ten change the first gap by between $-0.18$ and $0.09$
percent, and nine assortative rewirings in ten by between $-0.06$ and $0.20$ percent. Over the
completed transfer, $66$ percent of the disassortative rewirings and $30$ percent of the
assortative rewirings leave the first gap narrower than at the start. Both shares exceed the shares
at the start of the transfer in Table~\ref{tab:firstgap}. The rewirings whose direction reverses
along the path nearly all widen the first gap at the start and leave it narrower at the end
(Appendix~\ref{app:criterion}). For each of the first six gaps, the direction at the start of the
transfer is the direction of the net change in more than nine rewirings in ten, so the criterion
carries over to transfers of finite weight. Along the sequences of $40$ small transfers, the lower
bound of Corollary~\ref{cor:repeat} recovers a median of $90$ percent of the realized fall of the
first gap.\footnote{Since the criterion is exact, its sign agrees with a finite difference of the
gap in every test. When the transfers are small, the finite-mesh condition establishes contraction
at every point of the transfer for many of the gaps that contract, and the lower bound recovers at
least $75$ percent of the realized fall on 97 of the 100 networks (Appendix~\ref{app:criterion}).}

Along the trajectories of Section~\ref{sec:gapdist}, the small changes of single rewirings add
up.\footnote{The law of large numbers reconciles the shares of Table~\ref{tab:firstgap} with the end
states of Figure~\ref{fig:firstgap}. A single disassortative rewiring contracts the first gap in
three cases out of five and widens it in two, each time by a few hundredths of one percent, so over
a long sequence of such rewirings the first gap moves like a random walk with a small downward
drift. The drift grows in proportion to the number of rewirings, while the fluctuations around it
grow only with its square root, so after thousands of rewirings the sign of the net change is all
but certain although any one step is close to a toss-up. The assortative case is the mirror image,
with an upward drift.}
Figure~\ref{fig:firstgap}(a) and (b) present the first gap after the $30{,}000$ rewirings relative
to its initial value, on the 100 symmetric networks of $1{,}200$ vertices under each rule. With
lognormal weights, disassortative rewiring lowers the first gap to $0.83$ of its initial value on
average, to between $0.73$ and $0.95$ on nine networks in ten, and on 98 of the 100 networks.
Assortative rewiring raises the first gap to $2.6$ times its initial value on average, and to
between $2.3$ and $2.8$ on nine networks in ten. On the binary networks, disassortative rewiring
lowers the first gap to $0.27$ of its initial value on average, to between $0.16$ and $0.40$ on nine
networks in ten, and on every network, and assortative rewiring raises it to $2.5$ times its initial
value.\footnote{Unselected rewiring raises the first gap to $1.3$ times its initial value on every
weighted network, since the growth of the support lowers the second eigenvalue faster than the
Perron root, and leaves it at $0.98$ of its initial value on the binary networks, whose support does
not grow. The same growth of the support offsets part of the contraction under disassortative
rewiring with lognormal weights, which is smaller than on the binary networks and does not last to
the end of the trajectory. The median first gap falls to $0.78$ of its initial value by the time the
coefficient $r_{io}$ reaches about $-0.2$ and recovers part of the fall over the remaining rewirings,
whereas on the binary networks the first gap is still falling at the end of the trajectory on 95 of
the 100 networks.} Measured against the coefficient $r_{io}$, the first gap changes by about 5
percent for every tenth by which the coefficient falls under disassortative rewiring with lognormal
weights, by about 30 percent for every tenth by which it rises under assortative rewiring, and by
about 33 and 15 percent on the binary networks.\footnote{Each figure is the median over the 100
networks of the logarithm of the ratio of the final to the initial first gap, divided by the change
in $r_{io}$ along the trajectory.}

On directed networks, whose eigenvalues need not be real, we follow the distance between the two
eigenvalues of the closest pairs in place of a gap, over a transfer of one twentieth of the largest
admissible weight on 100 networks of $600$ vertices. Disassortative rewiring closes these pairs no
more often than unselected or assortative rewiring does, in about $48$ percent of the configurations
under each selection, whereas the radial derivative at the start of the transfer gives the direction
of the net change for $98$ percent of them (Appendix~\ref{app:pairstest}).

\subsection{Gaps of length zero}\label{sec:origin}

\begin{figure}[h!]\centering
  \includegraphics[width=0.62\linewidth]{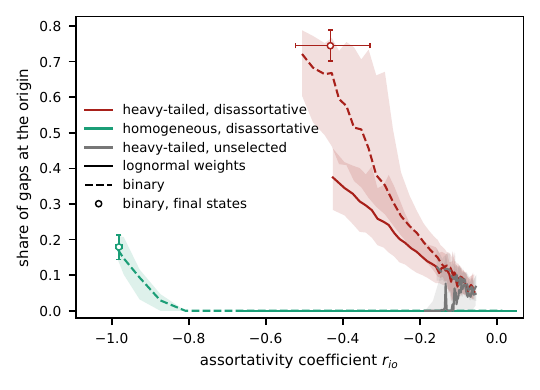}
  \caption{The share of adjacent gaps classified as zero against the assortativity
  coefficient $r_{io}$ on symmetric networks of $1{,}200$ vertices without self-loops. Lines give the
  median over the checkpoints of 100 networks in bins of the coefficient, with the range from the 5th
  to the 95th percentile shaded. Each open circle with bars gives the mean and the 5 to 95 percent
  range of the end states of the binary disassortative trajectories, which stall before $30{,}000$
  rewirings.
  }
  \label{fig:origin}
\end{figure}

We now measure how often rewiring makes eigenvalues coincide and so creates gaps of length
zero.\footnote{Since the criterion for the direction of a gap, Proposition~\ref{prop:criterion}, needs
distinct eigenvalues, these gaps lie outside its reach, whereas Proposition~\ref{prop:global} needs
no condition on the multiplicities.} The share of these gaps is the atom of the gap distribution at the origin. An eigenvalue of algebraic multiplicity
$m$ appears $m$ consecutive times in the ordered list and creates $m-1$ zero gaps, so
\begin{equation}
\nu_W(\{0\})
=
\frac{1}{n-1}
\sum_{\lambda}\bigl(m(\lambda)-1\bigr).
\label{eq:atom}
\end{equation}
Here the sum runs over the distinct eigenvalues $\lambda$ of $W$ of algebraic multiplicity
$m(\lambda)\geq2$. When the eigenvalue zero is repeated, it contributes $(m(0)-1)/(n-1)$, where
$m(0)$ is its algebraic multiplicity. For a symmetric $W$ this multiplicity equals the nullity
$\nullity W$.\footnote{For a directed or nonnormal matrix, $m(0)$ need not equal $\nullity W$, which
is the geometric multiplicity of zero. In general $\nullity W\leq m(0)$. The two agree when zero is
semisimple.}

We run six experiments on symmetric networks of $1{,}200$ vertices without self-loops, each
following 100 networks for up to $30{,}000$ rewirings. Four of the experiments are on heavy-tailed
networks and pair disassortative or unselected rewiring with weighted or binary edges. The other two are on homogeneous networks, which serve as a control for the role of the heavy tail.
A homogeneous network has degrees that follow a Poisson distribution, so that no vertex has many
more neighbors than another, and carries the same lognormal edge weights as the heavy-tailed
networks in the weighted case and unit weights in the binary case. The two experiments apply
disassortative rewiring to each. With continuous lognormal weights we do not expect two nonzero eigenvalues to coincide exactly. When
none do, the share of zero gaps on the weighted networks equals $\max\{\nullity W-1,0\}/(n-1)$. On
each of the 300 weighted trajectories the two agree at every checkpoint, and no gap classified as
zero lies away from the eigenvalue zero.

Figure~\ref{fig:origin} presents the results, with each line the median over the checkpoints of the
100 networks of one rule in bins of $r_{io}$. The binary disassortative trajectories stall before
$30{,}000$ rewirings, once almost no admissible rewiring is disassortative, so their end states
appear as open circles with bars.

The experiments show that disassortative rewiring creates gaps of length zero on heavy-tailed
networks, while unselected rewiring creates none. On heavy-tailed networks, weighted disassortative
rewiring raises the share of zero gaps from $0.08$ to $0.38$ on average as the mean $r_{io}$ falls
from $-0.09$ to $-0.38$. The final shares lie between $0.31$ and $0.46$ on nine networks in ten.
Weighted unselected rewiring instead removes the initial atom, within $2{,}600$ rewirings on the
median network and within $5{,}900$ rewirings on every network. On the homogeneous weighted networks
the share remains zero. The share of zero gaps at the end of the trajectory does not depend on the size of the network.
Under weighted disassortative rewiring it ends at $0.38$ on 20 networks of $4{,}800$ vertices and at
$0.37$ on 20 networks of $20{,}000$ vertices, and at the larger size the final shares of the 20
networks all lie between $0.35$ and $0.39$ (Appendix~\ref{app:design}). The binary experiments show that the rise in the share of
zero gaps does not come from the growth of the support.\footnote{On heavy-tailed networks,
unselected binary rewiring leaves the atom near $0.09$, whereas disassortative binary rewiring
raises it to $0.75$, and on homogeneous binary networks disassortative rewiring creates an atom of
$0.18$ once $r_{io}$ becomes sufficiently negative. Nothing in our analytical results explains why
disassortative rewiring creates the repeated zero eigenvalue. Since Proposition~\ref{prop:global}
needs no condition on the multiplicities, our results on the gaps taken together still hold on such
a spectrum, and each gap of length zero enters the spectral depth of a real spectrum as a zero
term.}

\section{Conclusion}\label{sec:conclusion}

We set out to say something, and to say it analytically, about how assortativity influences the
distribution of spectral gaps. We have proved that assortativity moves the gaps taken together
through one property, the Perron root, since the spectral depth, which is a weighted sum of these
gaps, moves with the Perron root. We have also shown that a disassortative rewiring lowers the Perron root, and with it
the spectral depth, whenever the Perron vectors rank the four rewired vertices as their strengths
would rank them. So a disassortative rewiring that lowers the Perron root narrows the gaps taken
together, whatever it does to any one gap. And we have derived an exact criterion for the direction in which an individual
gap moves at any point of a rewiring, together with a condition under which a gap contracts at every
point of the rewiring.

But we have not managed to derive claims that show how a particular spectral gap responds to
assortativity under the most general of circumstances. In fact, we have been able to show quite the
opposite. To say how a given gap responds to assortativity, one needs the counts of walks of every
length between the four rewired vertices, weighted by coefficients that depend on the whole spectrum
of the network or, approximately, on its leading eigenvalues. The counts of all but the longest of
these walks are compatible with both the contraction and the expansion of any gap. And so, perhaps,
the primary insight of this paper is a negative one. From the effect of a single rewiring on
assortativity, little can be said about what that rewiring does to how fast a network converges, a
rate that its first gap sets, or to the other dynamics that its individual gaps govern. In our
computational experiments, a sequence of disassortative rewirings tends to narrow the first gap on nearly every network,
and a sequence of assortative rewirings widens it. So assortativity does influence, through the first
spectral gap, the rate at which a network converges to its long-run state. The extent of that
influence depends on a variety of factors specific to the network in question, from the weights that
its edges carry to the walks that join the vertices being rewired.

\newpage
\bibliographystyle{plain}
\bibliography{gap_distribution}

\newpage
\appendix
\section{Auxiliary lemmas}\label{app:lemmas}

Throughout the appendices, the conjugate transpose of a matrix \(X\) is \(X^{*}\) and its nuclear
norm \(\lVert X\rVert_*\) is the sum of its singular values. The condition number of an invertible
matrix \(V\) is \(\kappa(V)=\lVert V\rVert_2\lVert V^{-1}\rVert_2\), and \(\delta_{ij}\) is the
Kronecker delta.

\begin{lemma}[Preservation of strengths and trace]\label{lem:preserve}
Along the transfer path \eqref{eq:path} and along the symmetric transfer, every row sum, every
column sum, every diagonal entry and the trace of $W(t)$ are constant.
\end{lemma}

\begin{proof}
Each of $x$ and $y$ has one entry equal to $+1$, one entry equal to $-1$, and all other entries
zero, so $x^{\mathsf T}\mathbf{1}=0$ and $y^{\mathsf T}\mathbf{1}=0$. Hence
$B\mathbf{1}=x\,(y^{\mathsf T}\mathbf{1})=\mathbf{0}$ and
$\mathbf{1}^{\mathsf T}B=(\mathbf{1}^{\mathsf T}x)\,y^{\mathsf T}=\mathbf{0}^{\mathsf T}$.
Multiplying by $t\delta$ and adding $W$ gives $W(t)\mathbf{1}=W\mathbf{1}$ and
$\mathbf{1}^{\mathsf T}W(t)=\mathbf{1}^{\mathsf T}W$ for every $t$. The $i$-th diagonal entry of $B$
is $x_iy_i$. The product $x_iy_i$ vanishes for every $i$ because $x$ is supported on $\{a,c\}$, $y$ is supported on
$\{d,b\}$, and the four vertices are distinct. Hence $\operatorname{diag}B=\mathbf{0}$ and
$\trace B=0$, so the diagonal and the trace of $W(t)$ are constant. The transpose
$B^{\mathsf T}=yx^{\mathsf T}$ also has zero row and column sums and a zero diagonal, and so does the
velocity $\delta(B+B^{\mathsf T})$ of the symmetric transfer.
\end{proof}

\begin{lemma}[Assortativity along the transfer path]\label{lem:assort}
Along the transfer path \eqref{eq:path}, the identity \eqref{eq:assortderiv} holds, and the
covariance $C_{io}(W(t))$ is affine in $t$,
\begin{equation}
C_{io}\bigl(W(t)\bigr)
=
C_{io}(W)
+
\frac{t\delta}{\omega}
\bigl(\din_a-\din_c\bigr)
\bigl(\dout_d-\dout_b\bigr),
\qquad 0\le t\le1 .
\label{eq:assortaffine}
\end{equation}
Under the symmetric transfer, the derivative of $C_{io}(W(t))$ is twice the
right-hand side of \eqref{eq:assortderiv}.
\end{lemma}

\begin{proof}
The four entries of $W(t)$ that move are
$w_{ab}-t\delta$, $w_{cd}-t\delta$, $w_{ad}+t\delta$ and $w_{cb}+t\delta$, so the total weight
$\omega$ is constant. By Lemma~\ref{lem:preserve} the vectors $\din$ and $\dout$ are constant,
and the row and column marginals of $\pi$ are $\dout/\omega$ and $\din/\omega$. Hence
$\bar d_{\mathrm{in}}=\sum_i(\dout_i/\omega)\din_i$ and
$\bar d_{\mathrm{out}}=\sum_j(\din_j/\omega)\dout_j$ are constant as well. Write
$u_i=\din_i-\bar d_{\mathrm{in}}$ and $v_j=\dout_j-\bar d_{\mathrm{out}}$, which are
constant in $t$.
Differentiating \eqref{eq:Cio} leaves only the derivative of $\pi$,
\[
\frac{d}{dt}C_{io}\bigl(W(t)\bigr)
=
\frac1\omega\sum_{ij}\frac{dw_{ij}}{dt}\,u_iv_j
=
\frac{\delta}{\omega}
\bigl(u_av_d+u_cv_b-u_av_b-u_cv_d\bigr).
\]
The bracket factors as $(u_a-u_c)(v_d-v_b)$, and the means cancel in each difference, which gives
\eqref{eq:assortderiv}. The right-hand side of \eqref{eq:assortderiv} does not depend on $t$, so
integrating from $0$ to $t$ gives \eqref{eq:assortaffine}.

Under the symmetric transfer the four transposed entries move as well, which keeps $\omega$
constant, and the velocity $\delta B^{\mathsf T}$ adds the bracket
$u_dv_a+u_bv_c-u_bv_a-u_dv_c=(u_d-u_b)(v_a-v_c)$. Because the network is symmetric, $\din=\dout$ and
the two means coincide. So $u=v$, this bracket equals $(u_a-u_c)(v_d-v_b)$, and the derivative
doubles.
\end{proof}

\begin{lemma}[Derivative of the Perron root]\label{lem:perron}
Under Assumption~\ref{ass:pf}, the Perron root and the unit positive left and right Perron vectors
$\ell(t)$ and $r(t)$ of $W(t)$ satisfy
\begin{equation}
\rho'(t)
=
\delta\,
\frac{
\bigl(\ell_a(t)-\ell_c(t)\bigr)
\bigl(r_d(t)-r_b(t)\bigr)
}{
\ell(t)^{\mathsf T}r(t)
}.
\label{eq:leftright}
\end{equation}
Under the symmetric transfer, $\ell(t)=r(t)$ and
$\rho'(t)=2\delta\bigl(r_a(t)-r_c(t)\bigr)\bigl(r_d(t)-r_b(t)\bigr)$.
\end{lemma}

\begin{proof}
Under Assumption~\ref{ass:pf} the Perron root is a simple eigenvalue, and the derivative formula
for a simple eigenvalue gives $\rho'=\ell^{\mathsf T}W'r/(\ell^{\mathsf T}r)$
\cite{kato,greenbaum}. Substituting $W'=\delta xy^{\mathsf T}$ and using
$\ell^{\mathsf T}x=\ell_a-\ell_c$ and $y^{\mathsf T}r=r_d-r_b$ gives
\eqref{eq:leftright}. For the symmetric transfer, $\ell=r$ with
$r^{\mathsf T}r=1$ and $W'=\delta(xy^{\mathsf T}+yx^{\mathsf T})$, so that
$\rho'=\delta\bigl((r^{\mathsf T}x)(y^{\mathsf T}r)+(r^{\mathsf T}y)(x^{\mathsf T}r)\bigr)
=2\delta(r_a-r_c)(r_d-r_b)$.
\end{proof}

\begin{lemma}[Eigenvalue velocities and moments]\label{lem:response}
Suppose that $W(t)$ has $n$ distinct eigenvalues $\lambda_1(t),\dots,\lambda_n(t)$, with spectral
projectors $P_1(t),\dots,P_n(t)$. Then the velocities $s_r(t)=\lambda_r'(t)$ satisfy
$s_r=\delta\,y^{\mathsf T}P_r\,x$ and $\sum_{r=1}^{n}s_r(t)=0$. The function $\mathcal{G}$ of
\eqref{eq:G} has its partial-fraction representation for every $z$ outside the spectrum and its
expansion at infinity for $\lvert z\rvert>\lVert W(t)\rVert_2$,
\begin{equation}
\mathcal{G}(z;t)
=
\sum_{r=1}^{n}\frac{s_r(t)}{z-\lambda_r(t)}
=
\sum_{q\ge 0}\frac{\mu_q(t)}{z^{q+1}},
\qquad
\mu_q(t)
=
\delta\,y^{\mathsf T}W(t)^qx
=
\sum_{r=1}^{n}\lambda_r(t)^q\,s_r(t).
\label{eq:reps}
\end{equation}
Under the symmetric transfer, the same identities hold with $2\delta$ in place of $\delta$.
\end{lemma}

\begin{proof}
The derivative formula for a simple eigenvalue gives $s_r=\trace(P_rW')$
\cite{kato,greenbaum}. Substituting $W'=\delta xy^{\mathsf T}$ and using cyclicity of the trace
gives $s_r=\delta\,y^{\mathsf T}P_r\,x$. The spectral resolution
$(zI-W)^{-1}=\sum_r(z-\lambda_r)^{-1}P_r$ inserted into \eqref{eq:G} gives the
partial-fraction representation. For $\lvert z\rvert>\lVert W\rVert_2$ the Neumann series
$(zI-W)^{-1}=\sum_{q\ge0}W^q/z^{q+1}$ converges and gives the expansion at infinity with
$\mu_q=\delta\,y^{\mathsf T}W^qx$. Expanding each matrix power as
$W^q=\sum_r\lambda_r^qP_r$ and using $s_r=\delta\,y^{\mathsf T}P_rx$ gives
$\mu_q=\sum_r\lambda_r^qs_r$. The eigenvalues sum to the trace, so
$\sum_rs_r=\frac{d}{dt}\trace W(t)=\trace W'=0$ by Lemma~\ref{lem:preserve}.

For the symmetric transfer, $W'=\delta(xy^{\mathsf T}+yx^{\mathsf T})$, and $P_r$ and $W^q$ are
symmetric. So $\trace(P_rW')=2\delta\,y^{\mathsf T}P_rx$ and $\trace(W^qW')=2\delta\,y^{\mathsf T}
W^qx$, and the identities hold with $2\delta$ in place of $\delta$.
\end{proof}

\begin{lemma}[Velocities from moments]\label{lem:recon}
Suppose that the eigenvalues of $W$ are distinct. Then the Vandermonde matrix
$\mathsf{V}(\lambda)$, with entries $\lambda_r^{q}$ for $0\le q\le n-1$ and $1\le r\le n$, is
invertible. The moment vector $\mu=(\mu_0,\dots,\mu_{n-1})^{\mathsf T}$ and the velocity vector
$s=(s_1,\dots,s_n)^{\mathsf T}$ satisfy $\mu=\mathsf{V}(\lambda)s$, so that
$s=\mathsf{V}(\lambda)^{-1}\mu$. The full Lagrange polynomial $L_r$ of the spectrum associated
with $\lambda_r$ satisfies $L_r(W)=P_r$ and $s_r=\delta\,y^{\mathsf T}L_r(W)\,x$, and its
coefficients in the monomial basis form the $r$-th row of $\mathsf{V}(\lambda)^{-1}$.
\end{lemma}

\begin{proof}
The moment identity of Lemma~\ref{lem:response} taken for $q=0,\dots,n-1$ is
$\mu=\mathsf{V}(\lambda)s$. The Vandermonde determinant is
$\prod_{i<j}(\lambda_j-\lambda_i)$, which is nonzero because the eigenvalues are distinct, so
$\mathsf{V}(\lambda)$ is invertible.

The polynomial $L_r(z)=\prod_{j\neq r}(z-\lambda_j)/(\lambda_r-\lambda_j)$ has degree $n-1$ and
satisfies $L_r(\lambda_j)=\delta_{rj}$, so the polynomial functional calculus gives
$L_r(W)=\sum_jL_r(\lambda_j)P_j=P_r$, and the identity $s_r=\delta\,y^{\mathsf T}P_rx$ of
Lemma~\ref{lem:response} becomes $s_r=\delta\,y^{\mathsf T}L_r(W)\,x$. In the monomial basis
$L_r(z)=\sum_{q=0}^{n-1}c_{rq}z^q$, and the interpolation conditions
$\sum_qc_{rq}\lambda_j^q=\delta_{rj}$ say that
$(c_{r0},\dots,c_{r,n-1})\,\mathsf{V}(\lambda)=e_r^{\mathsf T}$, so the coefficients of $L_r$ form
the $r$-th row of $\mathsf{V}(\lambda)^{-1}$.
\end{proof}

\begin{lemma}[Insufficiency of low-order moments]\label{lem:nonid}
Suppose that $\lambda_1>\cdots>\lambda_n$ are real and that $0\le M\le n-2$. Then for every
$s\in\R^{n}$ and every $i<n$ there are vectors $s^{+}$ and $s^{-}$ in $\R^{n}$ with the same
moments $\sum_r\lambda_r^qs_r$ as $s$ for every $0\le q\le M$ and with
\[
s^{+}_i-s^{+}_{i+1}>0>s^{-}_i-s^{-}_{i+1}.
\]
\end{lemma}

\begin{proof}
Partial fractions give $1/\prod_r(z-\lambda_r)=\sum_rh_r/(z-\lambda_r)$ with $h_r=\prod_{j\neq
r}(\lambda_r-\lambda_j)^{-1}$. The left side is $z^{-n}+O(z^{-n-1})$ at infinity, and expanding both
sides there gives
\[
\sum_r\lambda_r^qh_r=0
\quad(0\le q\le n-2),
\qquad
\sum_r\lambda_r^{n-1}h_r=1 .
\]
For every real $\tau$ and every $M\le n-2$, the vector $s+\tau h$ has the same moments as $s$
through order $M$. The $\lambda_r$ are decreasing, so $h_r$ alternates in sign with $r$, and
$h_i-h_{i+1}\neq0$ because $h_i$ and $h_{i+1}$ have opposite signs. The $i$-th gap velocity
along the line $\tau\mapsto s+\tau h$ is
\[
s_i-s_{i+1}+\tau(h_i-h_{i+1}).
\]
This velocity is affine in $\tau$ with the nonzero slope $h_i-h_{i+1}$, so it is positive at some
point $s^{+}$ of the line and negative at another point $s^{-}$.
\end{proof}

But in the exact criterion \eqref{eq:criterion} the coefficients of $Q_i$ in the monomial basis
are the differences of two rows of the inverse Vandermonde matrix of Lemma~\ref{lem:recon}, which
is ill-conditioned when the eigenvalues cluster. This ill-conditioning is the practical reason for
the two filters, whose errors we now bound.

\begin{lemma}[Filter error]\label{lem:filter}
Suppose that $W$ has $n$ distinct eigenvalues and spectral projectors $P_1,\dots,P_n$, that $V$ is
any invertible matrix such that $W=V\Lambda V^{-1}$, and that $f$ is a scalar function defined at
every eigenvalue, with $f(W)=Vf(\Lambda)V^{-1}$. Then for every matrix $E$ and every $r$,
\begin{equation}
\bigl\lvert\trace\bigl((f(W)-P_r)E\bigr)\bigr\rvert
\le
\lVert E\rVert_*\,\kappa(V)\,
\max_{1\le j\le n}
\lvert f(\lambda_j)-\delta_{rj}\rvert.
\label{eq:generalE}
\end{equation}
\end{lemma}

\begin{proof}
The spectral norm of $f(W)-P_r$ is at most
$\kappa(V)\max_j\lvert f(\lambda_j)-\delta_{rj}\rvert$, since
\[
f(W)-P_r
=
V\operatorname{diag}
\bigl(f(\lambda_j)-\delta_{rj}\bigr)_{j=1}^{n}
V^{-1}.
\]
Trace duality, $\lvert\trace(XY)\rvert\le\lVert X\rVert_2\lVert Y\rVert_*$ \cite{hornjohnson},
gives \eqref{eq:generalE}.
\end{proof}

We use a polynomial filter and a rational filter in the computations, and
\eqref{eq:generalE} gives the error of each once its values at the eigenvalues are bounded. For
the directed perturbation $E=\delta xy^{\mathsf T}$ of Definition~\ref{def:switch}, the nuclear
norm $\lVert E\rVert_*$ in \eqref{eq:generalE} is that of a rank-one matrix,
$\delta\lVert x\rVert_2\lVert y\rVert_2=2\delta$. For the perturbation
$E=\delta(xy^{\mathsf T}+yx^{\mathsf T})$ of the symmetric transfer, $E$ has the two nonzero
eigenvalues $\pm\delta\lVert x\rVert_2\lVert y\rVert_2=\pm2\delta$ because $x^{\mathsf T}y=0$, so
its nuclear norm is $4\delta$.

The polynomial filter applies to a real spectrum whose largest eigenvalues are separated from the
others, and it damps the remaining eigenvalues with a Chebyshev polynomial \cite{saad,shuman}.
Suppose that the retained eigenvalues satisfy $\lambda_1>\cdots>\lambda_R>\beta$ and that
$\lambda_{R+1},\dots,\lambda_n$ lie in $[\alpha,\beta]$, where $1\le R<n$ and $\alpha<\beta$. The
map $\varphi(z)=(2z-\alpha-\beta)/(\beta-\alpha)$ takes $[\alpha,\beta]$ onto $[-1,1]$ and each
retained eigenvalue $\lambda_r$ to $\xi_r=\varphi(\lambda_r)>1$. For $r\le R$, the Lagrange factor
of the retained eigenvalues is
\[
\Pi_r(z)
=
\prod_{\substack{j\le R\\j\neq r}}
\frac{z-\lambda_j}{\lambda_r-\lambda_j}.
\]
The polynomial filter $p_{r,m}$ multiplies this factor by a Chebyshev polynomial $T_m$ of the first
kind and degree $m\ge0$, scaled to equal one at $\lambda_r$. The number $\varepsilon_{r,m}$ bounds
the values of the filter at the eigenvalues that are not retained,
\begin{equation}
p_{r,m}(z)
=
\Pi_r(z)\,
\frac{T_m\bigl(\varphi(z)\bigr)}{T_m(\xi_r)},
\qquad
\varepsilon_{r,m}
=
\frac{
\max_{z\in[\alpha,\beta]}\lvert\Pi_r(z)\rvert
}{
\lvert T_m(\xi_r)\rvert
}.
\label{eq:cheb}
\end{equation}
The interpolation conditions give $p_{r,m}(\lambda_j)=\delta_{rj}$ for every $j\le R$. For $j>R$ the
eigenvalue $\lambda_j$ lies in $[\alpha,\beta]$, where $\lvert T_m(\varphi(z))\rvert\le1$, so
$\lvert p_{r,m}(\lambda_j)\rvert\le\varepsilon_{r,m}$. Every factor of $\Pi_r$ has the form
$(z-\lambda_j)/(\lambda_r-\lambda_j)$ with $\lambda_j>\beta$, so
$\lvert z-\lambda_j\rvert$ decreases on $[\alpha,\beta]$ and the maximum in
\eqref{eq:cheb} is attained at $\alpha$. Since $T_m(\xi)=\cosh(m\operatorname{arcosh}\xi)$ for
$\xi>1$ and $\cosh u\ge e^{u}/2$, Lemma~\ref{lem:filter} with $f=p_{r,m}$ shows that for the
directed perturbation the approximate velocity $\widehat{s}_r=\trace(p_{r,m}(W)W')$ satisfies
\begin{equation}
\lvert\widehat{s}_r-s_r\rvert
\le
2\delta\,\kappa(V)\,\varepsilon_{r,m},
\qquad
\varepsilon_{r,m}
\le
2\max_{z\in[\alpha,\beta]}\lvert\Pi_r(z)\rvert\,
e^{-m\operatorname{arcosh}\xi_r}.
\label{eq:filtbound}
\end{equation}
The symmetric transfer has the same bound with $4\delta$ in place of $2\delta$.

The rational filter applies also to a spectrum that is not real, and it needs only a circle
around the retained eigenvalue that excludes the others \cite{trefethen}. Suppose that a circle
with center $z_r$ and radius $\varrho_r>0$ encloses $\lambda_r$ and no other eigenvalue. The
rational filter with the $m\ge1$ equally spaced points $z_r+\varrho_re^{2\pi ik/m}$,
$k=0,\dots,m-1$, of that circle is
\begin{equation}
q_{r,m}(z)
=
\frac1m
\sum_{k=0}^{m-1}
\frac{\varrho_re^{2\pi ik/m}}{z_r+\varrho_re^{2\pi ik/m}-z}
=
\frac{1}{1-\bigl((z-z_r)/\varrho_r\bigr)^{m}} .
\label{eq:rational}
\end{equation}
The second expression in \eqref{eq:rational} follows by expanding each summand as a geometric series in
$e^{-2\pi ik/m}(z-z_r)/\varrho_r$, since averaging over $k$ annihilates every power that is not a
multiple of $m$ and leaves a geometric series in $((z-z_r)/\varrho_r)^m$. The two sides are
rational functions that coincide inside the circle, so they agree wherever both are defined. Suppose
that $\lvert\lambda_r-z_r\rvert/\varrho_r\le\varsigma<1$ and
$\min_{j\neq r}\lvert\lambda_j-z_r\rvert/\varrho_r\ge\varpi>1$, so that no quadrature point is an
eigenvalue. At the retained eigenvalue
$\lvert q_{r,m}(\lambda_r)-1\rvert\le\varsigma^m/(1-\varsigma^m)$, and at any other eigenvalue the
denominator in \eqref{eq:rational} is at least $\varpi^m-1$ in modulus, so
$\lvert q_{r,m}(\lambda_j)\rvert\le1/(\varpi^m-1)$. Lemma~\ref{lem:filter} with $f=q_{r,m}$ shows
that for the directed perturbation the approximate velocity $\widehat{s}_r=\trace(q_{r,m}(W)W')$
satisfies
\begin{equation}
\lvert\widehat{s}_r-s_r\rvert
\le
2\delta\,\kappa(V)\,\mathrm{err}_m,
\qquad
\mathrm{err}_m
=
\max\left\{
\frac{\varsigma^{m}}{1-\varsigma^{m}},
\frac{1}{\varpi^{m}-1}
\right\}.
\label{eq:ratbound}
\end{equation}
The symmetric transfer has the same bound with $4\delta$ in place of $2\delta$.

For a simple eigenvalue $\lambda_r(t)$ of $W(t)$, with right eigenvector $v_r(t)$ and left
eigenvector $w_r(t)$ normalized by $w_r^{*}v_r=1$, the spectral projector of $\lambda_r$ is
$P_r=v_rw_r^{*}$ and its condition number is
$\kappa_r(t)=\lVert w_r(t)\rVert_2\lVert v_r(t)\rVert_2$. The reduced resolvent of $\lambda_r$,
with the inverse taken on the range of $I-P_r$, is
\[
S_r(t)
=
\bigl(\lambda_r(t)I-W(t)\bigr)^{-1}(I-P_r).
\]
The largest distance between two branches $\lambda_i$ and $\lambda_j$ along the path is
$M_{ij}=\sup_{0\le t\le1}\lvert\lambda_i(t)-\lambda_j(t)\rvert$.

\begin{lemma}[Second derivatives of eigenvalues]\label{lem:curv}
Suppose that $E=W'$ is either the directed perturbation $\delta xy^{\mathsf T}$ or the perturbation
$\delta(xy^{\mathsf T}+yx^{\mathsf T})$ of the symmetric transfer, and that $\lambda_r(t)$ is a simple
eigenvalue of $W(t)$ for every $t\in[0,1]$. Then
\begin{equation}
\lambda_r''(t)
=
2\,w_r^{*}ES_rEv_r,
\qquad
\lvert\lambda_r''(t)\rvert
\le
A_r
:=
8\delta^2
\sup_{0\le t\le1}
\kappa_r(t)\,\lVert S_r(t)\rVert_2.
\label{eq:curv}
\end{equation}
The constant $A_r$ is finite. If $\lambda_i$ and $\lambda_{i+1}$ are simple on $[0,1]$, the
curvature constant $C_i=A_i+A_{i+1}$ bounds the second derivative of the $i$-th gap,
$\lvert g_i''(t)\rvert\le C_i$. If $\lambda_i$ and $\lambda_j$ are distinct and simple on
$[0,1]$, the curvature constant $F_{ij}$ of the pair bounds the second derivative of
$D_{ij}=\lvert\lambda_i-\lambda_j\rvert^2$,
\begin{equation}
\lvert D_{ij}''(t)\rvert
\le
F_{ij}
:=
8\delta^2
\sup_{0\le t\le1}
\bigl(\kappa_i(t)+\kappa_j(t)\bigr)^2
+
2M_{ij}\bigl(A_i+A_j\bigr).
\label{eq:paircurv}
\end{equation}
\end{lemma}

\begin{proof}
Since $\lambda_r$ is simple on $[0,1]$, the eigenvalue and its eigenvectors can be chosen analytic
in $t$ \cite{kato}, with the normalization $w_r^{*}v_r=1$ maintained along the path. Fix $t$ and
drop it from the notation.
Differentiating $(W-\lambda_r)v_r=0$ gives
$(\lambda_r-W)v_r'=(E-\lambda_r')v_r$, and the right-hand side lies in the range of $I-P_r$
because
\[
w_r^{*}(E-\lambda_r')v_r
=
\lambda_r'-\lambda_r'
=
0.
\]
Applying $S_r$ and using $S_r(\lambda_r-W)=I-P_r$ and $S_rv_r=0$ gives $(I-P_r)v_r'=S_rEv_r$. So
$v_r'=S_rEv_r+\alpha_1v_r$ for some scalar $\alpha_1$. The same
computation for the left eigenvector gives
$w_r^{*\prime}=w_r^{*}ES_r+\alpha_2w_r^{*}$ for some scalar $\alpha_2$, and differentiating
$w_r^{*}v_r=1$ gives $\alpha_1+\alpha_2=0$. Since
$\lambda_r'=w_r^{*}Ev_r$ and $E$ does not depend on $t$, differentiating once more gives the
identity in \eqref{eq:curv},
\[
\lambda_r''
=
w_r^{*\prime}Ev_r+w_r^{*}Ev_r'
=
2\,w_r^{*}ES_rEv_r
+
(\alpha_1+\alpha_2)\,w_r^{*}Ev_r
=
2\,w_r^{*}ES_rEv_r .
\]
The identity and $\kappa_r$ are both
invariant under $v_r\mapsto cv_r$, $w_r\mapsto w_r/\bar c$, so no further normalization
is needed.

The spectral norm is submultiplicative, so
\[
\lvert\lambda_r''\rvert
\le
2\lVert w_r\rVert_2\lVert E\rVert_2
\lVert S_r\rVert_2
\lVert E\rVert_2\lVert v_r\rVert_2
=
2\kappa_r\lVert S_r\rVert_2\lVert E\rVert_2^2.
\]
For either perturbation $\lVert E\rVert_2=2\delta$, which gives $A_r$. The eigenvectors and the reduced resolvent are continuous in $t$ while $\lambda_r$
stays simple, so the supremum over the compact interval is finite. The bound on $g_i''$ follows
from $g_i''=\lambda_i''-\lambda_{i+1}''$.

For a pair of branches, differentiating \eqref{eq:radial} gives
\[
D_{ij}''
=
2\lvert s_i-s_j\rvert^2
+
2\Real\Bigl[
\overline{\lambda_i-\lambda_j}\,
\bigl(\lambda_i''-\lambda_j''\bigr)
\Bigr].
\]
Since
$s_r=w_r^{*}Ev_r$ satisfies
$\lvert s_r\rvert\le\kappa_r\lVert E\rVert_2=2\delta\kappa_r$,
the first term is at most
$8\delta^2(\kappa_i+\kappa_j)^2$, and the second is at most $2M_{ij}(A_i+A_j)$.
\end{proof}

When $W(t)$ is diagonalizable on $[0,1]$, with $W(t)=V(t)\Lambda(t)V(t)^{-1}$, the constants of
Lemma~\ref{lem:curv} are bounded through the condition number of the whole eigenbasis. Whatever
the scaling
of the columns of $V$, the vector $v_r$ is a column of $V$ and $w_r^{*}$ a row of $V^{-1}$, so
$\kappa_r\le\lVert V\rVert_2\lVert V^{-1}\rVert_2=\kappa(V)$. The reduced resolvent is
$S_r=V\operatorname{diag}(d_1,\dots,d_n)V^{-1}$ with $d_j=(\lambda_r-\lambda_j)^{-1}$ for
$j\neq r$ and $d_r=0$, which gives
\[
\lVert S_r\rVert_2
\le
\frac{\kappa(V)}{\gamma_r},
\qquad
\gamma_r
=
\min_{j\neq r}
\lvert\lambda_r-\lambda_j\rvert.
\]
Hence $A_r\le8K^2\delta^2/\gamma_r$ with
$K=\sup_t\kappa(V(t))$ and $\gamma_r$ replaced by its infimum along the path, and
\[
F_{ij}
\le
32K^2\delta^2+2M_{ij}(A_i+A_j).
\]
These cruder constants are never smaller than the constants of Lemma~\ref{lem:curv}.

\section{Details of the computational experiments}\label{app:numerics}

We describe here the design of the computational experiments of Section~\ref{sec:computations}.
We also check the identities of Sections~\ref{sec:whole} and~\ref{sec:individual} and test the
sufficient conditions of those sections. The checks and the tests cover both the ordered gaps and
the pairs of complex eigenvalues.

\subsection{Design of the experiments}\label{app:design}

All experiments begin with configuration-model networks. In the heavy-tailed ensembles we draw the
in-degree and out-degree sequences from truncated power laws and couple them through a
Gaussian copula with a correlation we set to $0.7$. The homogeneous controls use Poisson degree sequences,
with mean $6$ except where stated. We discard self-loops and collapse parallel edges. Edge weights
are lognormal, the exponential of a normal variable with mean zero and standard deviation $0.5$.
When the diagonal is present, it is drawn from the same distribution. We restrict directed networks to their largest strongly connected
component. For the large directed trajectories we check connectivity at the start, at an
intermediate checkpoint and at the endpoint, and in every case the largest strongly connected
component still contains all $n$ vertices. For the symmetric experiments we forget orientation after pairing
the stubs, and restrict the resulting network to its largest connected component.
Since an independently weighted positive diagonal would make the matrix nonsingular almost surely and
remove the zero gaps being measured, we omit the diagonal in the experiments on gaps of length zero. Table~\ref{tab:numdesign} summarizes the design of each ensemble.

\begin{table}[h!]
\centering
\small
\caption{Design of the computational experiments. The power-law entry gives the exponent and the
degree range.}
\label{tab:numdesign}
\begin{tabular}{@{}>{\raggedright\arraybackslash}p{0.20\linewidth}>{\raggedright\arraybackslash}p{0.18\linewidth}>{\raggedright\arraybackslash}p{0.27\linewidth}>{\raggedright\arraybackslash}p{0.26\linewidth}@{}}
\toprule
Question & Matrix & Degree ensemble & Replication \\
\midrule
Spectral depth
& directed, $n=20{,}000$
& power law $2.5$, degrees $2$--$200$
& 100 networks per rule, about $200{,}000$ rewirings each, and 100 of $5{,}000$, 30 of
  $80{,}000$ and 10 of $320{,}000$ vertices \\
Gap scale and shape
& symmetric, $n=1{,}200$
& power law $2.5$, degrees $2$--$60$
& 100 networks per rule, $30{,}000$ rewirings each, and 30 of $2{,}400$ and 20 of $4{,}800$
  vertices \\
Individual gaps
& symmetric, $n=600$
& power law $2.2$, degrees $2$--$60$
& 100 networks with every admissible rewiring and $120$ completed transfers on each, three of the
  networks for the mesh test, and 30 each of $1{,}200$ to $4{,}800$ and 10 of $9{,}600$ vertices \\
Gaps of length zero
& symmetric, $n=1{,}200$
& power law $2.2$, degrees $1$--$200$; Poisson mean $6$
& 100 networks, up to $30{,}000$ rewirings each, and 20 to 30 of $2{,}400$ to $20{,}000$
  vertices \\
Pairs of complex eigenvalues
& directed, $n=600$
& power law $2.5$, degrees $2$--$60$
& 100 networks, five transfers per network under each of three selections \\
\bottomrule
\end{tabular}
\end{table}

Each trajectory is a sequence of accepted rewirings of Section~\ref{sec:switch}. Given donor
edges $a\to b$ and $c\to d$ whose crossed edges $a\to d$ and $c\to b$ are absent, the weighted trajectories transfer the largest admissible weight \(\delta=\min\{w_{ab},w_{cd}\}\). The directed
trajectories follow \eqref{eq:path}, and the symmetric trajectories follow the symmetric transfer
\(W(t)=W+t\delta(xy^{\mathsf T}+yx^{\mathsf T})\). Both transfers preserve every strength and the trace (Lemma~\ref{lem:preserve}). With continuous
lognormal weights the two donor edges generically differ in weight, so exactly one of them
disappears at the endpoint while the two crossed edges appear. Each accepted weighted rewiring
adds one edge to the support. Throughout the experiments, the admissible rewirings we draw or
enumerate are those whose crossed edges are absent at the start. The checkpoints at which we record
the spectrum number 41 on the symmetric trajectories and 51 on the directed ones, and successive
checkpoints are about a fifth further along the trajectory than the last.

We classify rewirings by the sign of the strength contrast $\chi_{io}$ of
Definition~\ref{def:contrast}, disassortative when $\chi_{io}<0$ and assortative when
$\chi_{io}>0$. And we compute both $\chi_{io}$ and $r_{io}$ from strengths that include the
diagonal. In the unselected control we accept admissible rewirings without regard to this sign, so
its trajectories add edges at the same rate as the selected weighted ones and control for the
growth of the support. In the binary control all off-diagonal weights equal one, and a completed
rewiring removes both donor edges while adding both crossed edges. The number of edges and every
degree are then preserved, and the support of a binary network does not grow along a trajectory.
So an effect of the selection that appears on binary networks cannot come from the growth of the
support.

The binary control of Section~\ref{sec:gapdist} keeps the diagonal of the weighted network and sets
every off-diagonal weight to one. For the spectral depth we also run every rule without a diagonal,
on the same networks and with the same stream of proposals. On the larger networks of the size
comparisons, the number of rewirings grows in proportion to the number of edges, so the trajectories
reach comparable values of $r_{io}$. On every symmetric network of $20{,}000$ vertices we count the
zero gaps by the deficiency of a maximum matching between the rows and the columns of the support,
which is the part of the nullity that the pattern of zeros alone forces, and which agrees with the
count from
the spectrum at $99$ percent of the checkpoints of the networks of $1{,}200$ vertices and falls short
by at most four elsewhere. A trajectory stops early when it uses up a budget of proposals, $200$ times
the target number of rewirings on directed networks and $400$ times on symmetric ones. Binary
disassortative trajectories without a diagonal stall in this way, near states in which almost no
admissible rewiring is disassortative.

Unless stated otherwise, we report the mean over independent network realizations, and shaded
regions and bars give the range from the 5th to the 95th percentile across networks. We reduce counts of pairs of complex eigenvalues to one representative of each conjugacy class when we summarize directions of motion. All spectral
computations use double-precision floating point. Integer counts are exact, whereas
statements that depend on computed eigenvalues carry floating-point error. We classify a gap as zero when it does not
exceed $2n\epsilon_{\mathrm{mach}}\max_r|\lambda_r|$, where $\epsilon_{\mathrm{mach}}$ is machine precision.

\subsection{Identities, the response of the Perron root and the ranking of the rewired vertices}\label{app:ranking}

Along the directed trajectories of Figure~\ref{fig:gapdist}(a) the largest deviation of any
strength from its initial value is $5.1\times10^{-13}$ with lognormal weights and zero on binary
networks. On a symmetric instance of 1{,}200 vertices the weighted sum $\sum_i(n-i)g_i$ of the gaps agrees with the spectral depth $n\rho-\trace W$, and the two sides of the first identity in \eqref{eq:anchor} agree at eight
values of $k$ from $1$ to $1000$. In both comparisons the relative residual is at most of order
$10^{-16}$. The left-hand
side $(n-k)g_k$ of the inequality in \eqref{eq:anchor} is 37 percent of $\mathcal{R}_k$ at $k=1$
and below 0.5 percent at $k=50$ and $k=1000$. On a directed network of 20{,}000 vertices the exact response of the Perron root of Lemma~\ref{lem:perron} agrees with a second-order finite difference of step
$0.02$ over 40 rewirings, with relative discrepancies whose median is of order $10^{-7}$ and whose
largest value is of order $10^{-5}$.

We test Corollary~\ref{cor:decline} on 40 disassortative rewirings of a heavy-tailed directed
network of 6{,}000 vertices and on 40 such rewirings of a homogeneous network with Poisson mean $8$.
On both networks the initial response $\rho'(0)$ of the Perron root is negative for about three
quarters of the rewirings (Table~\ref{tab:ranking}). In about as many rewirings the differences of
Perron coordinates $\ell_a-\ell_c$ and $r_d-r_b$ have the signs of the strength contrasts
$\chi_{\mathrm{in}}$ and $\chi_{\mathrm{out}}$. Since we make both comparisons at the initial matrices, they test the
hypothesis of Corollary~\ref{cor:decline}(i) only at the start of the transfer.

\begin{table}[h!]
\centering
\small
\caption{Corollary~\ref{cor:decline} at the initial matrices of 40 disassortative rewirings on each
of two directed networks. The first three rows count rewirings.}
\label{tab:ranking}
\begin{tabular}{@{}>{\raggedright\arraybackslash}p{0.38\linewidth}cc@{}}
\toprule
& Heavy-tailed, $n=6{,}000$ & Homogeneous, Poisson mean $8$\\
\midrule
$\rho'(0)<0$ & $31$ & $29$\\
Differences of Perron coordinates have the signs of the strength contrasts & $30$ & $29$\\
Norm-based condition of Corollary~\ref{cor:decline}(ii) met & $0$ & $0$\\
\addlinespace
At the two sources & & \\
\quad Coordinate-error bound $\sqrt{2}\lVert\ell-p\rVert_2$ & $0.622$ & $0.251$\\
\quad Median $\lvert\chi_{\mathrm{in}}\rvert$ & $0.00642$ & $0.00406$\\
\quad Largest $\lvert\chi_{\mathrm{in}}\rvert$ & $0.137$ & $0.0158$\\
\addlinespace
At the two targets & & \\
\quad Coordinate-error bound $\sqrt{2}\lVert r-q\rVert_2$ & $0.651$ & $0.248$\\
\quad Median $\lvert\chi_{\mathrm{out}}\rvert$ & $0.00947$ & $0.00274$\\
\quad Largest $\lvert\chi_{\mathrm{out}}\rvert$ & $0.229$ & $0.0137$\\
\bottomrule
\end{tabular}
\end{table}

None of the 80 rewirings meets the norm-based sufficient condition of
Corollary~\ref{cor:decline}(ii). The global coordinate-error bounds exceed even the largest strength
contrasts, by a factor of at least $2.8$ on the heavy-tailed network and of at least $15.8$ on the
homogeneous network.

Along the trajectories of Figure~\ref{fig:gapdist}(a) we also measure the ranking at every
checkpoint, on $60{,}000$ rewirings drawn uniformly from the current network. At the start, the response of the Perron root is negative for $69$ percent of the disassortative rewirings, and both differences
of Perron coordinates have the signs of the strength contrasts for $65$ percent, on networks of every
size from $5{,}000$ to $320{,}000$ vertices. By the end of the weighted disassortative trajectories,
the share with a negative response of the Perron root falls to $59$ percent. The norm-based condition of
Corollary~\ref{cor:decline}(ii) holds for none of the sampled rewirings at any checkpoint.

Table~\ref{tab:depthsize} reports the fall of the spectral depth along the disassortative
trajectories by network size. Without a diagonal, the fall rises slightly with size and its spread
narrows. With the lognormal diagonal, the Perron root ends at the largest diagonal entry on a growing
share of the networks, and these networks account for the smallest falls at every size. On the
networks of $20{,}000$ vertices, the two smallest weighted falls, $29.6$ and $33.5$ percent, and the
four smallest binary falls, from $20.0$ to $31.8$ percent, all end at this floor.

\begin{table}[h!]
\centering
\small
\caption{Fall of the spectral depth along the disassortative trajectories, in percent of its initial
value, mean over networks. The last column counts the networks on which the Perron root ends within
$0.1$ percent of the largest diagonal entry, for weighted and for binary networks.}
\label{tab:depthsize}
\begin{tabular}{@{}rrccccc@{}}
\toprule
& & \multicolumn{2}{c}{Lognormal diagonal} & \multicolumn{2}{c}{No diagonal} & \\
\cmidrule(lr){3-4}\cmidrule(lr){5-6}
Vertices & Networks & Weighted & Binary & Weighted & Binary & Floor binds\\
\midrule
$5{,}000$ & $100$ & $37.7$ & $35.7$ & $38.1$ & $37.5$ & $3$ and $7$\\
$20{,}000$ & $100$ & $39.0$ & $36.7$ & $39.6$ & $38.4$ & $3$ and $10$\\
$80{,}000$ & $30$ & $39.4$ & $36.1$ & $40.4$ & $38.8$ & $4$ and $6$\\
$320{,}000$ & $10$ & $36.2$ & $29.4$ & $40.4$ & $38.7$ & $7$ and $8$\\
\bottomrule
\end{tabular}
\end{table}

\subsection{The gap criterion and the finite-mesh condition}\label{app:criterion}

Except where we name another network, the computations of this subsection use a symmetric
heavy-tailed instance of 600 vertices and 3{,}253 edges with a lognormal diagonal, whose degrees
follow a power law of exponent 2.2 between 2 and 60. Its six largest eigenvalues are $25.68$,
$13.57$, $13.08$, $12.83$, $12.29$ and $12.00$. For the criterion test and the two filters we transfer one quarter of the largest admissible weight. Over 120 rewirings and the first six gaps, the sign predicted
by \eqref{eq:criterion} agrees with the sign of a central finite difference of the gap in all
720 tests. Under a transfer of the largest admissible weight, the sign of the initial velocity $g_i'(0)$ disagrees
with the sign of the net change $g_i(1)-g_i(0)$ in between $6.0$ and $7.6$ percent of the
rewirings for each of the first six gaps, over $120$ rewirings on each of the 100 networks of
Table~\ref{tab:firstgap}. On 30 networks of $1{,}200$ vertices the same share lies between $5.8$ and
$7.9$ percent. For the first gap the disagreements run almost entirely one way. Of the
disassortative rewirings on the networks of $600$ vertices, $5.5$ percent widen the first gap at
the start of the transfer and leave it narrower at the end, and $0.3$ percent do the reverse. For
the assortative rewirings the two shares are $5.3$ and $0.9$ percent.

The rational filter \eqref{eq:rational}, with each circle centered at a retained eigenvalue
$\lambda_r$ and of radius $\gamma_r/2$, recovers the velocities of the four
largest eigenvalues with 32 quadrature points to absolute errors between $5\times10^{-13}$ and
$6\times10^{-12}$. These errors lie below the bound
$4\delta\,\mathrm{err}_{32}=3.3\times10^{-10}$ that \eqref{eq:ratbound} gives for the symmetric transfer, for
which $\kappa(V)=1$. The polynomial filter \eqref{eq:cheb} reaches a
relative error of order $10^{-14}$ for the leading velocity by degree 40. For the other three
retained eigenvalues, whose values of $\xi_r$ lie between $1.04$ and $1.10$, the computed
relative error never falls below order one and exceeds $10^{18}$ at degree 80. The factor
$\Pi_r$ has to cancel retained eigenvalues whose Chebyshev values exceed that of $\lambda_r$ by a
multiple growing at least like $e^{0.91m}$, so rounding at those eigenvalues overwhelms the
exact-arithmetic bound \eqref{eq:filtbound}.

We test the finite-mesh condition \eqref{eq:mesh} with $J=8$ on three of the 100 symmetric
networks of Section~\ref{sec:individualnum}. Since $\lVert E\rVert_2=2\delta$, an eigenvalue moves by at most $\lVert E\rVert_2h/2=\delta h$ over
half a mesh step \cite{bauerfike}. The eigenvalues stay simple and ordered on each mesh interval when
disjoint disks with radii exceeding $\delta h$ surround the mesh eigenvalues. For a symmetric
matrix $\kappa_r=1$ and $\lVert S_r\rVert_2=1/\gamma_r$, so we obtain the curvature constant of
Lemma~\ref{lem:curv} directly from the distances $\gamma_r$. Between mesh points, Weyl's inequality lets us bound each distance $\gamma_r(t)$ below by its smallest
value at the mesh points less $2\delta h$. The share of the $2{,}400$ rewiring--gap tests over the first four gaps that meet the
resulting pathwise condition falls from $31.4$ percent at a transfer of a thousandth of the largest admissible weight to $0.2$ percent at a transfer of a tenth (Table~\ref{tab:mesh}). Using the
distances at the mesh points alone, without the Weyl bound between them, changes the outcome of at
most four tests at any transfer size. Since about $48$ percent of the tests have $g_i'<0$ at all
mesh points at every transfer, the fall in the share that meets the condition comes from the
curvature allowance. At a fixed matrix $g_i'$ grows linearly with $\delta$ and $C_i$ grows
quadratically. So the curvature allowance grows faster with the transfer than the gap velocity
does, and a fourfold finer mesh has the same first-order effect on the test as a fourfold smaller
transfer.

Over all 9{,}225{,}055 admissible rewirings of the 600-vertex instance, the share of rewirings
that contract the first $k$ gaps at once falls from $0.43$ at $k=1$ to $6.5\times10^{-7}$ at $k=8$
(Table~\ref{tab:joint}). No rewiring of this instance contracts the first nine gaps. Over the 100
networks of Table~\ref{tab:firstgap}, about a billion admissible rewirings in all, no rewiring
contracts the first nine gaps on 60 of the networks, and none contracts the first eleven on any of
them. Pooled over the 100 networks, the share conditional on the first $k-1$ gaps contracting falls
from $0.29$ at $k=2$ to $0.10$ at $k=8$. Independent signs would keep this conditional share near the unconditional share of about
one half. Since the velocities of the eigenvalues are the residues of one scalar function, the signs
of the individual gaps are far from independent.

\begin{table}[h!]
\centering
\small
\caption{The finite-mesh condition \eqref{eq:mesh} with $J=8$ over the $2{,}400$ rewiring--gap tests
on the first four gaps. The transfer is a share of the largest admissible weight. The last column counts
the tests on which the two versions of the condition disagree.}
\label{tab:mesh}
\begin{tabular}{@{}cccc@{}}
\toprule
& \multicolumn{2}{c}{Tests that meet the condition (percent)} & \\
\cmidrule(lr){2-3}
Transfer & Weyl bound between mesh points & Distances at mesh points alone & Tests that differ\\
\midrule
$10^{-3}$ & $31.4$ & $31.4$ & $1$\\
$3\times10^{-3}$ & $19.7$ & $19.7$ & $1$\\
$10^{-2}$ & $8.6$ & $8.8$ & $4$\\
$3\times10^{-2}$ & $2.4$ & $2.4$ & $0$\\
$10^{-1}$ & $0.2$ & $0.2$ & $0$\\
\bottomrule
\end{tabular}
\end{table}

\begin{table}[h!]
\centering
\small
\caption{Share of the 9{,}225{,}055 admissible rewirings of the 600-vertex instance that contract the
first $k$ gaps at once, and the median of the same share over the 100 networks of
Table~\ref{tab:firstgap}.}
\label{tab:joint}
\begin{tabular}{@{}lcccccccc@{}}
\toprule
$k$ & $1$ & $2$ & $3$ & $4$ & $5$ & $6$ & $7$ & $8$\\
\midrule
Share & $0.43$ & $0.11$ & $0.024$ & $0.0042$ & $5.6\times10^{-4}$ & $7.3\times10^{-5}$
& $6.0\times10^{-6}$ & $6.5\times10^{-7}$\\
Median & $0.43$ & $0.12$ & $0.026$ & $0.0044$ & $6.1\times10^{-4}$ & $6.7\times10^{-5}$
& $6.0\times10^{-6}$ & $5.2\times10^{-7}$\\
\bottomrule
\end{tabular}
\end{table}

Table~\ref{tab:firstgapsize} repeats the first experiment of Section~\ref{sec:individualnum} on larger
networks, with $20$ million rewirings drawn uniformly from the admissible ones of each network. The
share of disassortative rewirings with a negative response of the Perron root stays near $0.87$ at every size,
while the share of these rewirings that contract the first gap rises from $0.65$ to $0.80$. The inverse participation ratio of the Perron vector, the sum of the fourth powers of the entries of
the unit vector, whose reciprocal is the number of vertices over which the vector spreads, falls from
$0.014$ at $600$ vertices to $0.0014$ at $9{,}600$, whereas that of the second eigenvector stays near
$0.045$. So the Perron vector spreads over more vertices as the network grows while the second
eigenvector stays concentrated on about twenty, and the velocity of the second eigenvalue is small
for a growing share of rewirings, which is why the term from the Perron root weighs more on larger
networks. Along
the sequences of 40 transfers of Section~\ref{sec:individualnum}, every one of the 100 networks
reaches 40 accepted transfers, on average $30$ of the 40 are disassortative, and in every accepted transfer the realized fall of the first gap is at least its lower bound
$\zeta_{1,\tau}$. Over the 100 networks the bound recovers a median of $90$ percent of the realized fall, and at least
$75$ percent on 97 of them; on the instance above, the first gap falls from $12.115211$ to
$12.114973$, and the lower bound of $2.054\times10^{-4}$ is $86$ percent of the realized decrease
of $2.379\times10^{-4}$. When we restrict the proposals to disassortative rewirings, the lower
bound recovers a median of $91$ percent of the realized fall.

\begin{table}[h!]
\centering
\small
\caption{Share of admissible rewirings that contract the first gap at the start of the transfer, by
network size, mean over networks. The last two columns refer to the disassortative rewirings.}
\label{tab:firstgapsize}
\begin{tabular}{@{}rrccccc@{}}
\toprule
Vertices & Networks & Disassortative & Unselected & Assortative & $\rho'<0$ & Contract if $\rho'<0$\\
\midrule
$600$ & $100$ & $0.605$ & $0.435$ & $0.260$ & $0.874$ & $0.648$\\
$1{,}200$ & $30$ & $0.625$ & $0.443$ & $0.257$ & $0.870$ & $0.674$\\
$2{,}400$ & $30$ & $0.659$ & $0.453$ & $0.245$ & $0.871$ & $0.714$\\
$4{,}800$ & $30$ & $0.678$ & $0.459$ & $0.239$ & $0.869$ & $0.738$\\
$9{,}600$ & $10$ & $0.723$ & $0.474$ & $0.223$ & $0.870$ & $0.795$\\
\bottomrule
\end{tabular}
\end{table}

\subsection{Pairs of complex eigenvalues}\label{app:pairstest}

We now turn to directed networks, whose eigenvalues need not be real. Since the ordered gaps need a real spectrum, on these networks we follow the distance
between the two eigenvalues of a pair in place of a gap. We ask whether a pair closes or widens under a disassortative rewiring, and whether
its direction at the start of the transfer gives the direction of its net change. On each of
100 directed weighted networks of $600$ vertices, we draw five rewirings under each of three
selections, disassortative, assortative and unselected, and transfer one twentieth of the largest
admissible weight. For each transfer we follow the $60$ closest pairs of complex eigenvalues at the
initial matrix by nearest-neighbor continuation along the transfer. As a baseline we also follow
60 pairs that are not chosen for closeness, each joining a complex eigenvalue drawn at random to one
of its ten nearest complex neighbors.\footnote{Of the 1,500 transfers, 79 percent pass the
branch-tracking checks described below, and the shares we report change by at most
$0.002$ when we drop the others.}

Since a pair of eigenvalues of a real matrix and its conjugate pair have the same squared separation
and the same radial motion, we reduce the tracked pairs to one representative of each conjugacy
class. The reduction leaves $16{,}640$ configurations of closest pairs under each selection. No result of
Section~\ref{sec:individual} says whether a pair closes or widens under a disassortative
rewiring. Nor do we find such a regularity in the computations. Between the original matrix and the
rewired matrix, $48.6$ percent of the configurations close under disassortative rewiring, $48.4$ per
cent under unselected rewiring and $48.5$ percent under assortative rewiring. The differences from
unselected rewiring are $0.2$ and $0.1$ percentage points, with 95 percent intervals of about
$\pm1.4$ points from a bootstrap that resamples networks and then transfers within each network.
The shortfall below one half goes with closeness. Among the closest fifth of these pairs, measured
by their separation relative to the local mean spacing, between $46$ and $47$ percent close under
each selection, whereas about half of the baseline pairs close under each selection. The sign of the
initial radial derivative nevertheless agrees with the direction of the net change for $98.3$ per
cent of the configurations.

The absolute radial share of a pair,
\[
\frac{\lvert\Real[\overline{z_{ij}}(s_i-s_j)]\rvert}
     {\lvert z_{ij}\rvert\lvert s_i-s_j\rvert},
\]
is the absolute cosine of the angle between the separation $z_{ij}$ and the relative velocity
$s_i-s_j$. The share is one for purely radial motion and zero for pure rotation. The median radial share over the configurations is $0.82$, and the radial share is below $0.5$ for
$28$ percent of the configurations, under each of the three selections.
Although the initial radial derivative gives the direction of the net change of a pair in almost
every case, a substantial minority of the closest pairs move mainly by rotation.

At the transfer size used here, none of the closest pairs that close meets the
finite-mesh condition of Proposition~\ref{prop:certified}(ii). The condition asks the contraction rate
of a pair at each mesh point to exceed the curvature allowance $F_{ij}h/2$. Here $h$ is
the mesh size, and the curvature constant $F_{ij}$ of Lemma~\ref{lem:curv} bounds the second
derivative of the squared distance $D_{ij}$ along the path. So $F_{ij}h/2$ bounds how far the contraction
rate of the pair can change between mesh points. Since $F_{ij}$ takes suprema over the whole path of the
condition numbers $\kappa_r$ and the reduced-resolvent norms $\lVert S_r\rVert_2$ of the two
eigenvalues, which are costly to bound, we test the condition with a lower bound on $F_{ij}$. The
lower bound takes these suprema over the mesh points only and replaces $\lVert S_r\rVert_2$ by
$1/\gamma_r$. Here $\gamma_r$ is the distance from $\lambda_r$ to the rest of the spectrum. Since a
maximum over the mesh points cannot exceed the supremum over the path, and
$\lVert S_r\rVert_2\ge1/\gamma_r$ for every simple eigenvalue, the true $F_{ij}$ is at least as large
as the lower bound and can only make the condition harder to meet. On 30 of the networks, even with
the lower bound the curvature allowance $F_{ij}h/2$ exceeds the smallest contraction rate at the
evaluation points by a factor of at least $9$ and of about $3{,}300$ at the median. The condition numbers of the two eigenvalues drive this failure, since with $\kappa_r=1$, the value
of a normal matrix, the lower bound would no longer rule out the condition for about a tenth of
these pairs. The lower bound controls the relative velocity of a pair through the norm of the rewiring. So the
bound does not shrink with the separation $\lvert z_{ij}\rvert$ of the pair. The radial derivative
$D_{ij}'$ is at most $2\lvert z_{ij}\rvert\lvert s_i-s_j\rvert$ in magnitude, and the pairs we follow
are the closest in the spectrum. So their radial derivatives are small. To establish that these
pairs close at every point of a transfer, the finite-mesh condition would need a curvature constant
built from the velocities of the two eigenvalues themselves, in place of a bound through the norm of
the rewiring.

We evaluate the transfers above at 81 points along the path and follow the
branches by nearest-neighbor assignment. A transfer passes the tracking checks when the assignment is
bijective at every step and each step is smaller than half the smaller of the eigenvalue spacings at
the two ends of the step. Of the 1{,}500 transfers, 1{,}173 pass both checks at 81 points. We
evaluate the others at 161 and then 321 points, which rescues only five more, since their failures
come from eigenvalues that pass close to one another rather than from a coarse mesh. On the
transfers that pass, the median relative discrepancy over the evaluation points between
\eqref{eq:radial} and a central finite difference is at most $1.5\times10^{-5}$, and the
corresponding discrepancy for the second-derivative formula in Lemma~\ref{lem:curv} is at most
$1.0\times10^{-3}$.

The separation of a pair relative to the local mean spacing divides the distance between its two
eigenvalues by the mean, over the two eigenvalues, of the average distance to their six nearest
neighbors. The closest pairs have relative separations between $0.02$ and $1.33$, and the baseline
pairs between $0.10$ and $3.27$.

We test the finite-mesh condition of Proposition~\ref{prop:certified}(ii) on the first 30 networks,
under all three selections and for all five transfers, 450 transfers in all. The lower bound on
$F_{ij}$ uses the condition numbers $\kappa_r(t_m)$ and the distances $\gamma_r(t_m)$ at every mesh
point $t_m$. Before conjugacy reduction, $13{,}036$ of the closest tracked pairs are closer together
in the rewired matrix than in the original matrix, and $12{,}530$ of them have $D_{ij}'<0$ at every
evaluation point. For each of these $12{,}530$ pairs, the finite-mesh condition fails already with the
lower bound. The ratio of the curvature allowance formed from the lower bound to the smallest
contraction rate at the evaluation points has minimum $9.3$ and median about $3{,}300$, and the
larger condition number of the two eigenvalues has median $33$. With $\kappa_r=1$ in place of the
computed condition numbers, the lower bound would no longer rule out the condition for $1{,}335$ of
these pairs. Among the $13{,}248$ closing baseline pairs that contract at every evaluation point,
four meet the condition with the lower bound, which leaves open whether they meet it with the true
$F_{ij}$. On a subsample of $8{,}100$ eigenvalue branches, the exact reduced-resolvent norm
$\lVert S_r\rVert_2$ exceeds $1/\gamma_r$ by a median factor of $41$, and by a factor between $19$ and
$135$ on nine branches in ten.

\section{Notation}\label{app:notation}

Table~\ref{tab:notation} lists the symbols of the paper, grouped by topic in the order in which
the topics appear, with the place where each symbol is introduced. Symbols used only inside a
single proof are left out.

{\small
\begin{longtable}{@{}>{\raggedright\arraybackslash}p{0.19\linewidth}>{\raggedright\arraybackslash}p{0.60\linewidth}>{\raggedright\arraybackslash}p{0.15\linewidth}@{}}
\caption{Notation.}\label{tab:notation}\\
\toprule
Symbol & Meaning & Introduced in\\
\midrule
\endfirsthead
\toprule
Symbol & Meaning & Introduced in\\
\midrule
\endhead
\bottomrule
\endfoot
\bottomrule
\endlastfoot
\multicolumn{3}{@{}l}{General notation}\\*
\cmidrule{1-3}
$e_i$ & $i$-th standard basis vector & Section~\ref{sec:switch}\\
$E_{ij}$ & matrix $e_ie_j^{\mathsf T}$, whose only nonzero entry is a one in position $(i,j)$
& Section~\ref{sec:switch}\\
$\mathbf{1}$ & vector of ones & Section~\ref{sec:switch}\\
$X^{\mathsf T}$, $X^{*}$ & transpose and conjugate transpose of $X$
& Section~\ref{sec:switch}, Appendix~\ref{app:lemmas}\\
$\trace X$, $\spec X$ & trace and spectrum of $X$ & Section~\ref{sec:switch}\\
$\nullity X$ & nullity of $X$, the dimension of its null space & Section~\ref{sec:switch}\\
$\operatorname{diag}(v)$, $\operatorname{diag}X$ & diagonal matrix with the entries of $v$ on its
diagonal, and the diagonal of $X$ & Section~\ref{sec:switch}\\
$\lVert v\rVert_2$, $\lVert X\rVert_2$ & Euclidean norm of a vector and spectral norm of a matrix
& Section~\ref{sec:switch}\\
$\lVert X\rVert_*$ & nuclear norm, the sum of the singular values of $X$ & Appendix~\ref{app:lemmas}\\
$\kappa(V)$ & condition number $\lVert V\rVert_2\lVert V^{-1}\rVert_2$ of an invertible matrix
& Appendix~\ref{app:lemmas}\\
$\delta_u$, $\delta_{ij}$ & unit point mass at $u$, and the Kronecker delta
& Section~\ref{sec:switch}, Appendix~\ref{app:lemmas}\\
$I$ & identity matrix & \eqref{eq:G}\\
$\angle(u,v)$ & angle between the vectors $u$ and $v$ & Section~\ref{sec:perron}\\
$\Real z$, $\overline{z}$ & real part and complex conjugate of $z$ & Section~\ref{sec:whole},
\eqref{eq:radial}\\
$\epsilon_{\mathrm{mach}}$ & machine precision & Appendix~\ref{app:design}\\
\addlinespace
\multicolumn{3}{@{}l}{The rewiring}\\*
\cmidrule{1-3}
$W=(w_{ij})$ & weight matrix of a directed network, with $w_{ij}\ge0$ the weight of the edge from
$i$ to $j$ & Section~\ref{sec:switch}\\
$n$ & number of vertices & Section~\ref{sec:switch}\\
$a,b,c,d$ & the four distinct rewired vertices, the edges $a\to b$ and $c\to d$ giving weight to
the crossed edges $a\to d$ and $c\to b$ & Section~\ref{sec:switch}\\
$x$, $y$ & $e_a-e_c$ and $e_d-e_b$ & Definition~\ref{def:switch}\\
$B$ & direction $xy^{\mathsf T}=E_{ad}+E_{cb}-E_{ab}-E_{cd}$ of the rewiring
& Definition~\ref{def:switch}\\
$\delta$ & transferred weight, with $0<\delta\le\min\{w_{ab},w_{cd}\}$ & Definition~\ref{def:switch}\\
$t$ & fraction of the weight already transferred, $0\le t\le1$ & \eqref{eq:path}\\
$W(t)$ & transfer path $W+t\delta B$, and $W+t\delta(xy^{\mathsf T}+yx^{\mathsf T})$ under the
symmetric transfer & \eqref{eq:path}, Section~\ref{sec:switch}\\
$W'$ & velocity of the path, $\delta B$ for the directed transfer and
$\delta(xy^{\mathsf T}+yx^{\mathsf T})$ for the symmetric transfer & Section~\ref{sec:switch}\\
$E$ & the velocity $W'$ in the proof of Proposition~\ref{prop:criterion} and in
Lemma~\ref{lem:curv}, and any matrix in Lemma~\ref{lem:filter} & Proposition~\ref{prop:criterion}\\
$\rho(t)$ & Perron root of $W(t)$, its spectral radius & Section~\ref{sec:switch}\\
\addlinespace
\multicolumn{3}{@{}l}{Assortativity}\\*
\cmidrule{1-3}
$\dout$, $\din$ & out-strength and in-strength vectors $W\mathbf{1}$ and $W^{\mathsf T}\mathbf{1}$
& \eqref{eq:Cio}\\
$\omega$, $\pi_{ij}$ & total weight $\sum_{ij}w_{ij}$, and the share $w_{ij}/\omega$ of the edge
from $i$ to $j$ & \eqref{eq:Cio}\\
$C_{io}(W)$ & covariance between the in-strength at the source of an edge and the out-strength at
its target, with edges weighted by $\pi_{ij}$ & \eqref{eq:Cio}\\
$\bar d_{\mathrm{in}}$, $\bar d_{\mathrm{out}}$ & edge-weighted means of source in-strength and
target out-strength & \eqref{eq:Cio}\\
$r_{io}$ & normalized coefficient $C_{io}/(\sigma_{\mathrm{in}}\sigma_{\mathrm{out}})$
& Section~\ref{sec:computations}\\
$\sigma_{\mathrm{in}}$, $\sigma_{\mathrm{out}}$ & edge-weighted standard deviations of source
in-strength and target out-strength & Section~\ref{sec:computations}\\
\addlinespace
\multicolumn{3}{@{}l}{Gaps and the spectral depth}\\*
\cmidrule{1-3}
$\lambda_1,\dots,\lambda_n$ & eigenvalues, repeated by algebraic multiplicity and, for a real
spectrum, in decreasing order; along a path, $\lambda_r(t)$ is an eigenvalue branch
& Definition~\ref{def:gaps}\\
$g_i$ & adjacent gap $\lambda_i-\lambda_{i+1}$ & \eqref{eq:gaps}\\
$\nu_W$ & gap distribution $(n-1)^{-1}\sum_{i}\delta_{g_i}$ & Definition~\ref{def:gaps}\\
$\mathcal{R}(W)$ & spectral depth $n\rho-\trace W$, the total distance of the eigenvalues beneath
the Perron root along the real axis; for a real spectrum, the weighted sum $\sum_i(n-i)g_i$ of the gaps & Definition~\ref{def:depth}, \eqref{eq:Rtrace}\\
$u$ & threshold on the size of a gap in the tail bound & \eqref{eq:tail}\\
$\mathcal{R}_k$ & spectral depth $\sum_{r\ge k}(\lambda_k-\lambda_r)$ beneath $\lambda_k$
& \eqref{eq:Rk}\\
$\Delta\lambda_r$, $\Delta\mathcal{R}$ & changes $\lambda_r(1)-\lambda_r(0)$ and
$\mathcal{R}(W(1))-\mathcal{R}(W(0))$ from the original matrix $W(0)$ to the rewired matrix $W(1)$
& Corollary~\ref{cor:anchor}\\
$I_+$ & indices of the gaps that are positive at $t=0$ & \eqref{eq:depthshare}\\
$\phi_i$ & proportional contraction $(g_i(0)-g_i(1))/g_i(0)$ of the $i$-th gap
& \eqref{eq:depthshare}\\
$\psi_i$ & share $(n-i)g_i(0)/\mathcal{R}(W(0))$ of the initial depth carried by the $i$-th gap
& \eqref{eq:depthshare}\\
$\mathcal{D}$ & proportional fall $1-\mathcal{R}(W(1))/\mathcal{R}(W(0))$ in the spectral depth
& \eqref{eq:depthshare}\\
$Z$ & weighted size $\sum_{i\notin I_+}(n-i)g_i(1)$ at the end of the gaps that start at zero & \eqref{eq:depthshare}\\
$\theta$ & threshold on the proportional contraction & Corollary~\ref{cor:proportional}\\
$m(\lambda)$ & algebraic multiplicity of the eigenvalue $\lambda$ & \eqref{eq:atom}\\
$\nu_W(\{0\})$ & share of gaps equal to zero, the atom of the gap distribution at the origin
& \eqref{eq:atom}\\
\addlinespace
\multicolumn{3}{@{}l}{The response of the Perron root}\\*
\cmidrule{1-3}
$\ell(t)$, $r(t)$ & unit positive left and right Perron vectors of $W(t)$, equal under the
symmetric transfer & Section~\ref{sec:perron}\\
$p$, $q$ & normalized strength vectors $\din/\lVert\din\rVert_2$ and $\dout/\lVert\dout\rVert_2$
& Definition~\ref{def:contrast}\\
$\chi_{\mathrm{in}}$, $\chi_{\mathrm{out}}$, $\chi_{io}$ & source, target and combined strength
contrasts $p_a-p_c$, $q_d-q_b$ and $\chi_{\mathrm{in}}\chi_{\mathrm{out}}$; a rewiring is
disassortative when $\chi_{io}<0$ & \eqref{eq:chi}\\
$\varepsilon_{\mathrm{in}}(t)$, $\varepsilon_{\mathrm{out}}(t)$ & coordinate errors, the parts of
$\ell_a-\ell_c$ and $r_d-r_b$ that the strength contrasts leave out & \eqref{eq:eps}\\
$\eta(t)$ & inner product $\ell(t)^{\mathsf T}r(t)$ of the Perron vectors & \eqref{eq:perronfactor}\\
$\bar\varepsilon_{\mathrm{in}}$, $\bar\varepsilon_{\mathrm{out}}$ & bounds on the coordinate errors
along the transfer & \eqref{eq:uniform}\\
\addlinespace
\multicolumn{3}{@{}l}{Stochastic networks}\\*
\cmidrule{1-3}
$\mathcal{P}$, $\mathcal{P}(t)$ & weight matrix $\operatorname{diag}(\dout)^{-1}W$ of the stochastic network, and the
same matrix along the path & Section~\ref{sec:whole}\\
\addlinespace
\multicolumn{3}{@{}l}{Velocities and moments}\\*
\cmidrule{1-3}
$s_r(t)$, $s$ & velocity $\lambda_r'(t)$ of the $r$-th eigenvalue, and the vector of velocities
& Section~\ref{sec:individual}\\
$P_r$ & spectral projector of $\lambda_r$ & Section~\ref{sec:individual}, Lemma~\ref{lem:response}\\
$\mathcal{G}(z;t)$ & matrix element $\delta\,y^{\mathsf T}(zI-W(t))^{-1}x$ of the resolvent, whose
residues are the velocities & \eqref{eq:G}\\
$\mu_q(t)$, $\mu$ & moment $\delta\,y^{\mathsf T}W(t)^qx$, a signed count of weighted walks of
length $q$ from the targets to the sources, and the vector $(\mu_0,\dots,\mu_{n-1})$
& Section~\ref{sec:individual}, Lemma~\ref{lem:recon}\\
$\mathsf{V}(\lambda)$ & Vandermonde matrix with entries $\lambda_r^q$
& Section~\ref{sec:individual}, Lemma~\ref{lem:recon}\\
$L_r$ & full Lagrange polynomial of the spectrum associated with $\lambda_r$, with $L_r(W)=P_r$
& Section~\ref{sec:individual}, Lemma~\ref{lem:recon}\\
$Q_i$ & polynomial $L_i-L_{i+1}$ of the criterion for the $i$-th gap & \eqref{eq:criterion}\\
$t_0$, $M$ & point of the path, and order of the first moment that need not vanish
& Proposition~\ref{prop:patterns}\\
\addlinespace
\multicolumn{3}{@{}l}{Pairs of complex eigenvalues}\\*
\cmidrule{1-3}
$z_{ij}(t)$ & difference $\lambda_i(t)-\lambda_j(t)$ of two eigenvalue branches & \eqref{eq:radial}\\
$D_{ij}(t)$ & squared distance $\lvert z_{ij}(t)\rvert^2$ between the two branches
& \eqref{eq:radial}\\
\addlinespace
\multicolumn{3}{@{}l}{The finite-mesh condition}\\*
\cmidrule{1-3}
$J$, $h$, $t_m$ & number of mesh intervals, mesh width $1/J$, and mesh points $mh$
& Proposition~\ref{prop:certified}\\
$\overline v_{i,m}$, $\overline d_{ij,m}$ & upper bounds on $g_i'(t_m)$ and $D_{ij}'(t_m)$ at the
mesh points & Proposition~\ref{prop:certified}\\
$v_r(t)$, $w_r(t)$ & right and left eigenvectors of $\lambda_r(t)$, normalized by $w_r^{*}v_r=1$;
the left eigenvector $w_r$ is distinct from the edge weights $w_{ij}$
& Lemma~\ref{lem:curv}\\
$S_r(t)$ & reduced resolvent $(\lambda_rI-W)^{-1}(I-P_r)$ of $\lambda_r$ & Lemma~\ref{lem:curv}\\
$\kappa_r(t)$ & condition number $\lVert w_r\rVert_2\lVert v_r\rVert_2$ of $\lambda_r$
& Lemma~\ref{lem:curv}\\
$A_r$ & bound on $\lvert\lambda_r''\rvert$ along the path & \eqref{eq:curv}\\
$C_i$, $F_{ij}$ & curvature constants bounding $\lvert g_i''\rvert$ and $\lvert D_{ij}''\rvert$;
$C_ih/2$ and $F_{ij}h/2$ are the curvature allowances & Lemma~\ref{lem:curv}\\
$M_{ij}$ & largest distance $\sup_t\lvert\lambda_i(t)-\lambda_j(t)\rvert$ between two branches
& Lemma~\ref{lem:curv}\\
$\gamma_r$ & distance $\min_{j\ne r}\lvert\lambda_r-\lambda_j\rvert$ from $\lambda_r$ to the rest
of the spectrum & Section~\ref{sec:mesh}, Appendix~\ref{app:lemmas}\\
$K$ & largest condition number $\sup_t\kappa(V(t))$ of an eigenbasis along the path
& Appendix~\ref{app:lemmas}\\
$W^{(\tau)}$, $\delta^{(\tau)}$, $B^{(\tau)}$, $N$ & matrix, transferred weight and direction of the $\tau$-th
rewiring in a sequence of $N$ rewirings, the superscript $(\tau)$ marking any quantity of that
rewiring & Corollary~\ref{cor:repeat}\\
$\zeta_{i,\tau}$ & lower bound on the rate at which the $i$-th gap contracts during the $\tau$-th
rewiring & Corollary~\ref{cor:repeat}\\
\addlinespace
\multicolumn{3}{@{}l}{Filters}\\*
\cmidrule{1-3}
$V$, $\Lambda$ & matrix of eigenvectors and diagonal matrix of eigenvalues, with
$W=V\Lambda V^{-1}$ & Lemma~\ref{lem:filter}\\
$f$ & scalar function defined at every eigenvalue, with $f(W)=Vf(\Lambda)V^{-1}$
& Lemma~\ref{lem:filter}\\
$R$ & number of retained eigenvalues in the polynomial filter & \eqref{eq:cheb}\\
$[\alpha,\beta]$ & interval that contains the eigenvalues that are not retained & \eqref{eq:cheb}\\
$\varphi$, $\xi_r$ & affine map of $[\alpha,\beta]$ onto $[-1,1]$, and its value
$\varphi(\lambda_r)$ at a retained eigenvalue & \eqref{eq:cheb}\\
$\Pi_r$ & Lagrange factor over the retained eigenvalues & \eqref{eq:cheb}\\
$T_m$ & Chebyshev polynomial of the first kind of degree $m$ & \eqref{eq:cheb}\\
$p_{r,m}$, $\varepsilon_{r,m}$ & polynomial filter for $\lambda_r$, and its bound on the
eigenvalues that are not retained & \eqref{eq:cheb}\\
$\widehat s_r$ & velocity recovered by a filter & Section~\ref{sec:individual}, \eqref{eq:filtbound}\\
$z_r$, $\varrho_r$ & center and radius of a circle that encloses $\lambda_r$ and no other
eigenvalue & \eqref{eq:rational}\\
$q_{r,m}$ & rational filter for $\lambda_r$ with $m$ quadrature points & \eqref{eq:rational}\\
$\varsigma$, $\varpi$ & bounds on the distances of $\lambda_r$ and of the other eigenvalues from
$z_r$, relative to $\varrho_r$ & \eqref{eq:ratbound}\\
$\mathrm{err}_m$ & error of the rational filter at the eigenvalues & \eqref{eq:ratbound}\\
\addlinespace
\multicolumn{3}{@{}l}{Indices}\\*
\cmidrule{1-3}
$i$, $j$, $k$, $r$ & indices of eigenvalues and gaps; in Proposition~\ref{prop:certified}, $k$ is
also the number of leading gaps tested together; the index $r$ is distinct from the right Perron
vector $r(t)$ and from the coefficient $r_{io}$ & throughout\\
$q$ & order of a moment, distinct from the normalized out-strength vector $q$
& throughout\\*
$m$ & index of a mesh point, degree of the polynomial filter, and number of quadrature points of
the rational filter & throughout\\
\end{longtable}
}

\end{document}